\documentclass[12pt]{iopart}

\UseRawInputEncoding
\usepackage{iopams}
\usepackage{amssymb,amsfonts,euscript,graphicx,bm,axodraw2}

\usepackage{fancyhdr}            % Permits header customization. See header section below.
\fancypagestyle{plain}{
    \lhead{}
    \fancyhead[R]{\thepage}
    \fancyhead[L]{}
    
    \fancyfoot{}
}

\begin{document}

%\begin{frontmatter}

\journal{Journal of Physics: Condensed Matter}



\title{Spatial non-locality of electronic correlations beyond GW approximation}

\author{A.~L.~Kutepov}
\address{Condensed Matter Physics and Materials Science Department, Brookhaven National Laboratory, Upton, NY 11973}
\ead{akutepov@bnl.gov}

\begin{abstract}
The question of spatial locality of electronic correlations beyond GW approximation is one of the central issues of the famous combination of GW and dynamical mean field theory, GW+DMFT. In this work, the above question is addressed directly (for the first time) by performing calculations with and without assumption of locality of the corresponding diagrams. For this purpose we use sc(GW+G3W2) approach where the higher order part (G3W2) is evaluated with fully momentum dependent Green's function G and screened interaction W and with "local" variant, where the single site approximation is assumed for both G and W. For all three materials studied in this work (NiO, $\alpha$-Ce, LiFeAs), we have found the spatial non-locality effects to be strong. For NiO and LiFeAs they, in fact, are decisive for the proper evaluation of vertex corrections. The results of this study have direct impact on our understanding of approximations made in practical implementations of GW+DMFT method, where all diagrams beyond GW (DMFT part) are assumed to be local. Taking into account the fact that the first diagrams beyond GW represent the most important contribution also in GW+DMFT calculations, we conclude that the basic assumption of GW+DMFT, namely the locality of diagrams evaluated in the DMFT part, is not as good as it is believed to be.
\end{abstract}

\vspace{2pc}
\noindent{\it Keywords}: GW approximation; Vertex correction; GW+DMFT

\submitto{\JPCM}
%\maketitle

\section{Introduction}
\label{intr}

Combination of GW approximation \cite{pr_139_A796} and dynamical mean field theory \cite{rmp_68_13} (DMFT) which is commonly abbreviated as GW+DMFT was proposed almost twenty years ago by S. Biermann et al., \cite{prl_90_086402}. It is an appealing combination of two "worlds": GW approach which features large basis sets and \textbf{k} (momentum) dependent functions and DMFT with its minimal basis sets and local (single site) description. It is based on an intuitive assumption that nonlocal effects can be taken care of by low order diagrams (GW part) whereas higher order diagrams are essentially local in space and can be well described by DMFT. Despite its attractive features, GW+DMFT was implemented for realistic materials only a few years ago, \cite{prm_1_043803}, and its applications still are very sparse, \cite{prl_90_086402,prb_94_201106,prr_2_013191,prx_10_041047}. There are, however, numerous studies which address certain aspects of the approach \cite{prl_77_3391,prl_109_237010,prb_61_5184,prb_63_115110,prb_91_121107,prb_95_115107,prb_90_165138,
prb_103_045121,prb_95_245130,prb_85_035115}. The sparsity of applications could be explained by the fact that the method is computationally expensive. However, there might be another reason, namely that the above mentioned assumption about sufficiency of GW to describe all non-local effects is not justified well enough.

The interest in spatial non-locality effects in relation to DMFT calculations is raising recently. The issue was addressed by different variants of multisite extensions, \cite{prb_75_045118,prb_95_115107,prb_61_12739,prb_94_205110,prb_79_045133,prl_118_106404}, as well as by studying specific to GW+DMFT questions \cite{philmagb_82_1413,prl_96_226403,prl_123_256401,prb_99_245156,prb_101_174509,prb_102_035109,prb_103_045121,prb_103_155107}. As the majority of DMFT-related applications are still done using a combination of density functional theory (DFT) and DMFT, the interest is not limited to higher order diagrams beyond GW but rather consists in non-locality effects which are not captured by the DFT+DMFT. In this case, non-locality effects inherent to the GW itself also contribute, so the error related to their neglect is generally larger than in GW+DMFT case. Looking at the details of the studies on this subject, one can notice that practically all of them are done along the same line. Namely, one assumes a certain ansatz, such as the single site approximation for the DMFT part in DFT+DMFT calculations, and by direct comparison of the results with the experimental ones one decides about importance (or non-importance) of the effects which are not captured by the assumed ansatz. Undoubtedly, this approach reveals strong non-locality effects when DFT+DMFT is not able to describe experimental properties. However, more interesting and also more dangerous situation appears when DFT+DMFT does reproduce some experimental features. The danger of this situation consists in the fact that it is very suggestive for the following conclusion: DFT+DMFT captures all important physics of the studied material. In fact, it can also be that the experimental results are reproduced because of parameters (such as Hubbard U/J and the way the double counting in DFT+DMFT is handled) adjustment. That is, the real physics behind the measured property could be quite different from the one the DFT+DMFT suggests. Clearly, one should use more direct way to assess the role of non-locality effects in order to exclude ambiguities.

The non-locality issue in GW+DMFT, which is the main object of this study, is even more delicate than in DFT+DMFT. On the one hand, there is no adjustable parameters in GW+DMFT, at least in its original formulation, \cite{prl_90_086402}. Also, undoubtedly, considerable part of non-locality effects is taken care of by the GW part. On the other hand, certain issues with practical implementation of GW+DMFT for crystalline materials, such as local approximation (on intermediate steps) for the Coulomb (or screened Coulomb) interaction and polarizability bring some ambiguities in the GW+DMFT as well. The reason is that above mentioned functions are intrinsically nonlocal (long ranged) and their replacement by local functions is rather questionable.  In this respect, recent study of the non-local effects in LiFeAs \cite{prb_103_155107} where authors use the difference between experimental spectra and QSGW (quasiparticle self-consistent GW approach) spectra as a measure of higher order effects is interesting. However, the authors' principal conclusion that the above difference can be described assuming locality of self energy (beyond QSGW) doesn't answer the question whether this higher order self energy (even if it is indeed local) can be quantitatively correctly evaluated using local approximation at the intermediate steps. In other words, exact self energy could be a result of mutual cancellation of a number of contributions (including the non-local ones). From this point of view, the question whether exact self energy is local or non-local is of secondary importance. The main question should, instead, be formulated as the following: Is it possible to obtain accurate results for higher order self energy assuming local approximation for all intermediate steps of the evaluation? This question is directly related to GW+DMFT where self energy is evaluated as a sum (infinite) of all diagrams build from local Green's function and local screened interaction.

This study addresses the question of non-locality effects in the diagrams beyond GW using direct method. In it, the same diagrams are evaluated in two different ways: with and without assumption of locality of functions involved in the evaluation of the diagrams. For simplicity, we consider only the next order of diagrams (as compared to the GW approximation) for both polarizability and self energy. For a number of weakly correlated materials this kind of vertex correction already represents a very good approximation \cite{prb_94_155101,prb_95_195120,prb_96_035108}. For materials with stronger correlations, first order vertex correction is not sufficient but, even for such materials, these diagrams represent a major contribution and their accurate evaluation is decisive. Clearly, the effect of non-locality for these diagrams, if important, can tell us a lot about its total influence also in the case of strongly correlated materials. Thus, even if we do not evaluate all diagrams beyond GW, as it is supposedly done (for local diagrams) in GW+DMFT, we unambiguously address the issue of non-locality in GW+DMFT method.

\section*{Methods and calculation setups}\label{meth}

\begin{figure}[t]
\begin{center}\begin{axopicture}(200,56)(0,0)
\SetPFont{Arial-bold}{28}
\SetWidth{0.8}
\Text(10,10)[l]{$\Psi$ =}
\Text(35,10)[l]{$-\frac{1}{2}$}
\GCirc(75,10){20}{1}
\Photon(55,10)(95,10){2}{5.5}
\Text(110,10)[l]{+}
\Text(128,10)[l]{$\frac{1}{4}$}
\GCirc(160,10){20}{1}
\Photon(160,-10)(160,30){2}{5.5}
\Photon(140,10)(180,10){2}{5.5}
\end{axopicture}
\end{center}
\caption{Diagrammatic representation of $\Psi$-functional which includes the simplest non-trivial vertex. First diagram on the right hand side stands for scGW approximation, whereas total expression corresponds to sc(GW+G3W2) approximation.}
\label{diag_Psi}
\end{figure}

\begin{figure}[t]
\begin{center}\begin{axopicture}(200,56)(0,0)
\SetPFont{Arial-bold}{28}
\SetWidth{0.8}
\Text(10,10)[l]{$P$  =}
\Photon(48,10)(55,10){2}{2.5}
\GCirc(75,10){20}{1}
\Photon(95,10)(102,10){2}{2.5}
\Text(120,10)[l]{$-$}
\Photon(143,10)(150,10){2}{2.5}
\GCirc(170,10){20}{1}
\Photon(190,10)(197,10){2}{2.5}
\Photon(170,-10)(170,30){2}{5.5}
\end{axopicture}
\end{center}
\caption{Diagrammatic representation of irreducible polarizability in the simplest vertex corrected scheme sc(GW+G3W2).}
\label{diag_P}
\end{figure}

\begin{figure}[t]
\begin{center}\begin{axopicture}(200,56)(0,0)
\SetPFont{Arial-bold}{28}
\SetWidth{0.8}
\Text(10,10)[l]{$\Sigma$ = -}
\Line(38,10)(112,10)
\PhotonArc(75,10)(30,0,180){2}{8.5}
\Text(120,10)[l]{$+$}
\Line(133,10)(207,10)
\PhotonArc(160,10)(20,0,180){2}{6.5}
\PhotonArc(180,10)(20,180,360){2}{6.5}
\end{axopicture}
\end{center}
\caption{Diagrammatic representation of self energy in the simplest vertex corrected scheme sc(GW+G3W2).}
\label{diag_S}
\end{figure}

\begin{figure}[t]
\begin{center}       
\includegraphics[width=8.5 cm]{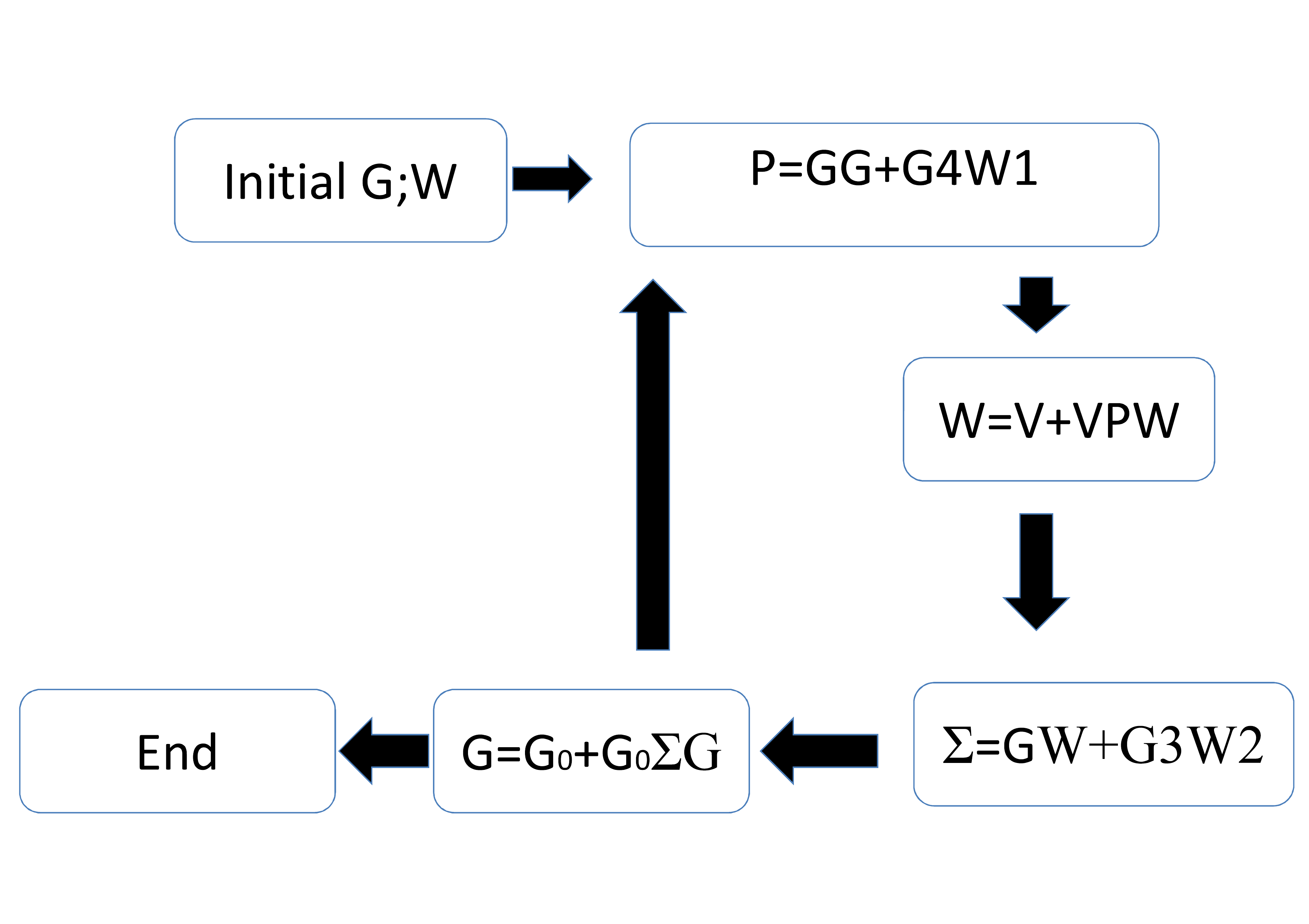}
\caption{Flowchart of sc(GW+G3W2) calculation. In the expressions for polarizability and self energy, the first terms on the right hand side represent the parts which are used in scGW approach, whereas the second terms are vertex corrections.}
\label{flow}
\end{center}
\end{figure}

All calculations in this work were performed using code FlapwMBPT \cite{flapwmbpt}. In this study we use scGW method to generate initial Green's function G and screened interaction W for subsequent vertex corrected iterations. Principal results of this study are obtained with sc(GW+G3W2) approximation, \cite{prb_94_155101,arx_2105_03770}. This self-consistent vertex corrected scheme was successfully applied to study electronic structure of simple (mostly sp) materials \cite{prb_95_195120} some time ago and, recently, was used to evaluate the band gap of van der Waals ferromagnet CrI$_{3}$ \cite{arx_2105_07798} and electronic structure of LaNiO$_{2}$ and CaCuO$_{2}$ \cite{arx_2105_03770}. Both, scGW and sc(GW+G3W2), are based on L. Hedin's theory \cite{pr_139_A796}. They can also be defined using $\Psi$-functional formalism of Almbladh et al. \cite{ijmpb_13_535}. Corresponding $\Psi$-functional which includes vertex corrections is shown in Fig. \ref{diag_Psi}. In Fig. \ref{diag_Psi}, the first diagram corresponds to scGW approximation, whereas the sum of the first and the second diagram represents sc(GW+G3W2) approximation. Diagrammatic representations for irreducible polarizability (Fig. \ref{diag_P}) and for self energy (Fig. \ref{diag_S}) in scGW and in sc(GW+G3W2) follow from the chosen approximation for $\Psi$-functional. Technical details of the GW part were described in Refs. \cite{prb_85_155129,cpc_219_407}. Detailed account of the algorithms for sc(GW+G3W2) and also for other vertex corrected schemes implemented in the FlapwMBPT code can be found in Refs. \cite{prb_94_155101,prb_95_195120,prb_96_035108,arx_2105_03770}. Briefly, the algorithms are demonstrated in Fig. \ref{flow} where flowchart of sc(GW+G3W2) methods is shown. In scGW calculations, all steps are the same but polarizability and self energy are evaluated without vertex corrections, i.e. without second parts on the right hand side of the corresponding blocks of the chart.

Let us now specify the setup parameters used in the calculations. Principal structural parameters for the studied solids have been collected in Table \ref{list_s}. The most important set up parameters have been collected in Table \ref{setup_s}. All calculations have been performed for the electronic temperature $600K$. For NiO, we assumed the ferromagnetic (FM) ordering for simplicity instead of experimentally observed antiferromagnetic (AFM) ordering.

\begin{table}[t]
\caption{Structural parameters of the solids studied in this work. Lattice parameters are in Angstroms, MT radii are in atomic units (1 Bohr radius), and atomic positions are given relative to the three primitive translation vectors. Experimental structural data from Refs. \cite{jacs_55_527,cgd_10_4428} are used.} \label{list_s}
\begin{center}
\begin{tabular}{@{}c c c c c c} &Space&&&Wyckoff&$R_{MT}$\\
Solid &group&a(\AA)&c(\AA)&positions&(a$_{B}$)\\
\hline\hline
NiO&225 &4.1684 & &Ni: 0;0;0&Ni: 2.19\\
& & &&O: 1/2;1/2;1/2  &O: 1.69\\
Ce&225 &4.824 && 0;0;0& 3.22\\
LiFeAs&129 &3.7678 &6.3151&Li: 1/4;1/4;0.625&Li: 2.72\\
& & &&Fe: -1/4;1/4;0  &Fe: 2.27\\
& & &&As: 1/4;1/4;0.23626  &As: 2.27\\
\end{tabular}
\end{center}
\end{table}

\begin{table}[b]
\caption{Principal setup parameters for the studied solids. The following abbreviations are introduced: $\Psi$ is for wave functions, $\rho$ is for the electronic density, and $V$ is for Kohn-Sham potential. $N_{bnd}$ is the number of bands used to expand Green's function and self energy in GW/Vertex part of the calculations.} \label{setup_s}
\begin{center}
\begin{tabular}{@{}c c c c c c} &&$L_{max}$&&\textbf{k}-mesh &N$_{bnd}$\\
Solid &Semicore&$\Psi/\rho,V$&$RK_{max}$&GW/Vertex &GW/Vertex \\
\hline\hline
NiO&Ni: 3s,3p&6/6&8.0 &6*6*6/2*2*2 &100/28\\
& O: 2s&5/5&&&  \\
$\alpha$-Ce& 5s,5p,4d&5/5&7.0 &6*6*6/2*2*2 &70/25\\
LiFeAs&Li: 1s&6/6&8.0 &6*6*4/2*2*2 &200/47\\
& Fe: 3s,3p&6/6&&&  \\
& As: 3s,3p,3d&6/6&&&  \\
\end{tabular}
\end{center}
\end{table}

In order to perform vertex corrected calculations using single site (local) approximation we used local orbitals confined inside their muffin-tin spheres as a basis set (instead of \textbf{k}-dependent band states in full calculations). This approximation was applied to the vertex part only, whereas in GW part we always use full \textbf{k}-dependent basis functions. For clarity, let us say a bit more about local approximation. For screened interaction W, the implementation of this approximation is relatively straightforward. The reason is that W is expanded in the so called product basis (PB) set (see for instance Ref. \cite{cpc_219_407} for specifics of definitions in the FlapwMBPT code). Product basis set, by construction, differentiates between muffin-tin spheres and the interstitial region. Therefore, local approximation is simply implemented leaving in the expansion only those pairs of PB functions (representing two space arguments of W) which belong to the same muffin-tin sphere. Situation with local approximation is slightly more involved and described in \ref{locappr}. Basis set of local orbitals (see \ref{locappr}) included both the solutions of radial equations ($\phi$) and their energy derivatives ($\dot{\phi}$) for specific set of orbitals: Ni(4s,3d); O(2s,3s,2p); Ce(6s,6p,5d,4f); Li(2s,2p); Fe(4s,3d); As(5s,4p). Below, we will use appreviation 'local' to distinguish this kind of calculations as opposed to the full \textbf{k}-dependent calculation which will be abbreviated as 'full'. In order to mimic minimal basis sets usually applied in DMFT calculations, the second variant of local approximation was also studied. In this second variant, only some of the above orbitals (and also on specific atoms) were used: Ni(3d) in NiO; Ce 4f in $\alpha$-Ce, and Fe(3d) in LiFeAs. For this (minimal basis set) calculations we will use abbreviation 'locmin', or, if we discuss a specific material, more concise abbreviation like 'Ni 3d only'. In DMFT calculations, normally only one orbital for every angular momentum channel is used. But it can be a Wannier function or other specially designed for single site calculations function. We do not specifically optimize the basis functions for local calculations but the use of double basis set ($\phi$ and $\dot{\phi}$) should compensate for that. 

One important clarification has to be given however. Whereas in "full" calculations the results depend only very marginally on such setup parameters as muffin-tin radius, specific choice of the shape of $\phi$ and $\dot{\phi}$, and the number of bands included in the expansion (\ref{G_1}), the situation with single site approximation is more complicated. In local approximations, G and W "operate" in a certain subspace which effectively is cut out from the full space of basis functions. As a result, results do depend noticeably on the choice of muffin-tin radius, specific choice of the shape of $\phi$ and $\dot{\phi}$, and on the number of bands included in the expansion (\ref{G_5}). This dependence is similar to the dependence of DFT+DMFT results on projectors and Wannier functions discussed recently in Ref. \cite{prb_103_195101}. With NdNiO${2}$ as a model system, and using different Wannier and projector methods to define the correlation problem, authors of Ref. \cite{prb_103_195101} have found different results for the orbital and band mass enhancements and for the orbital occupancies. They also have found different implications regarding the importance of multiorbital effects and charge transfer physics. Another finding is that the use of interaction parameters derived from the constrained random phase approximation only enhances the difference in results. Clearly, if one knows the answer from the beginning, like from full space evaluation of diagrams (in this study) or from experiment, one can "adjust" the way of construction of the "local" subspace in order to minimize the deviation from already known final answer. If, however, the final answer is unknown the above dependence represents a problem. Studying all these dependencies on the choice of "local" subspace is, however, beyond the scope of this work. In order to make the results of this work obtained within "local" approximations well defined, they were all obtained with fixed values of muffin-tin radii (Table \ref{list_s}), with $\phi$ and $\dot{\phi}$ as obtained from DFT radial equations with linearization centers chosen as centers of mass of occupied band, and with the restriction of bands included in summation in Eq. (\ref{G_5}) specified in Table \ref{setup_s}.

\section{Results}
\label{res}

\subsection{NiO}
\label{nio}

\begin{figure}[t] 
\begin{center}       
    \fbox{\includegraphics[width=6.5 cm]{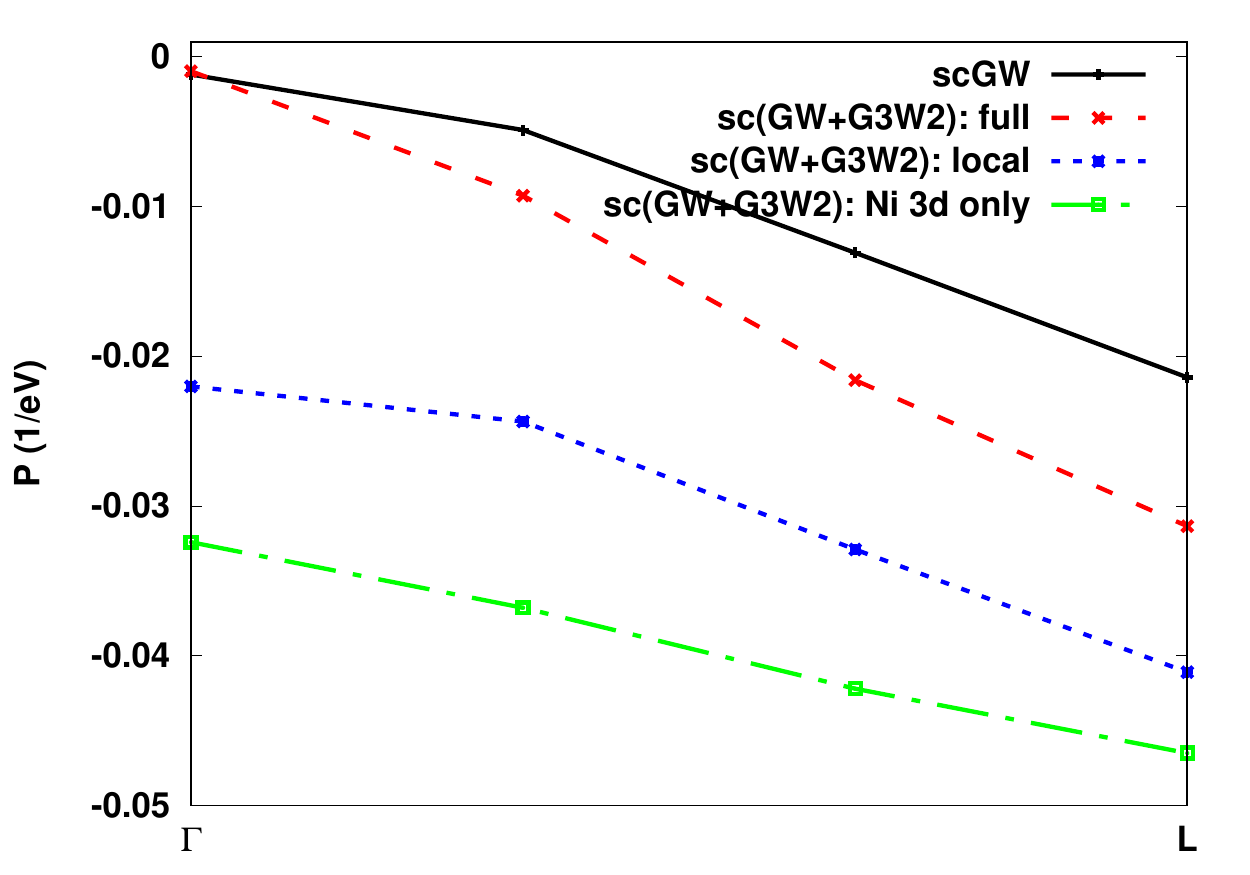}}   
    \hspace{0.02 cm}
    \fbox{\includegraphics[width=6.5 cm]{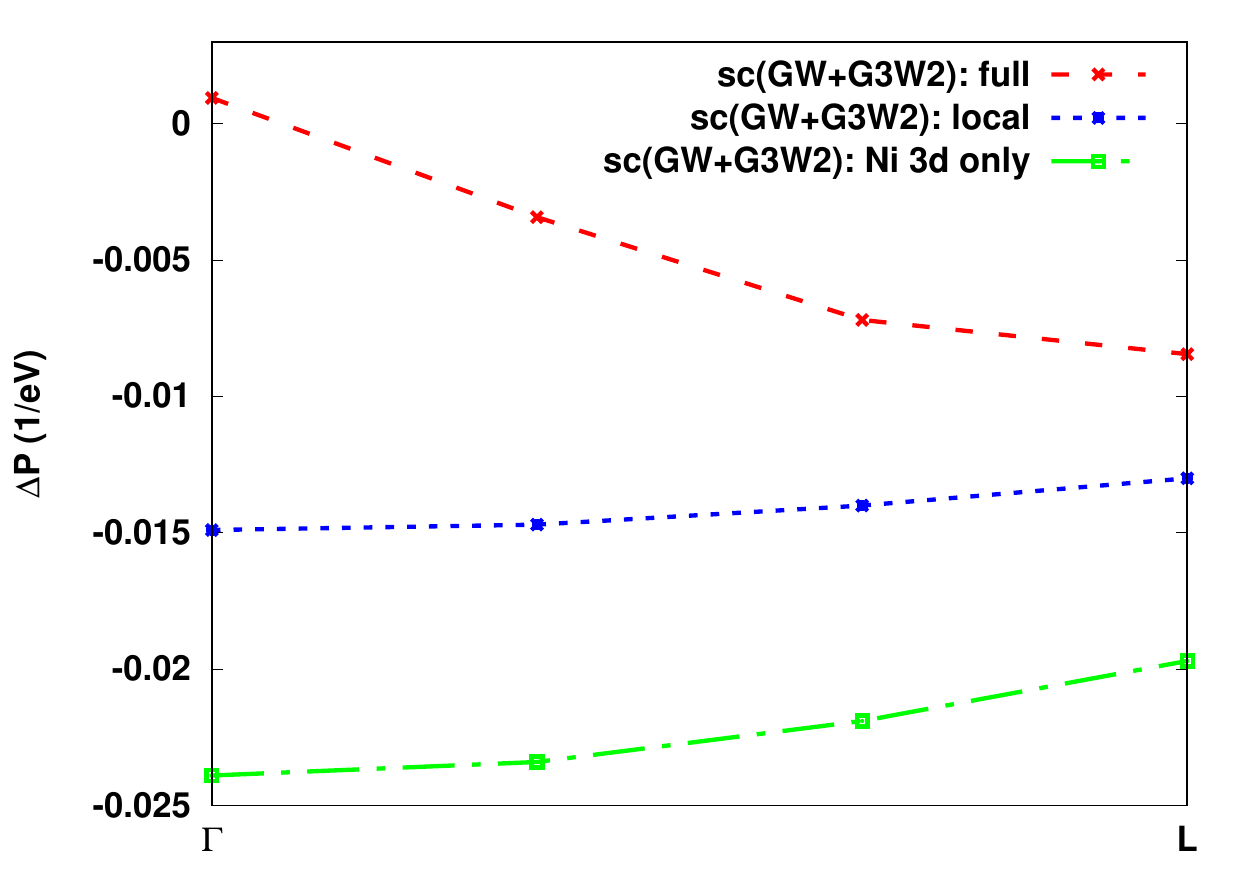}}  
    \hspace{0.02 cm} 
    \caption{Left window: The components $P^{\mathbf{k}}_{\mathbf{G}=\mathbf{G}'=0}(\nu=0)$ (see Eq. (\ref{head}) for precise definition) of calculated irreducible polarizability as functions of momentum $\mathbf{k}$ along the direction $\Gamma$-L in the Brillouin zone. In vertex corrected results, the difference (as compared to scGW) comes not only from vertex correction but also includes self-consistency effects in G and W. Right window: The same but only vertex correction ($\Delta P$) to polarizability (instead of full polarizability) is compared.}
    \label{p_nio}
\end{center}
\end{figure}

Principal physical quantity for comparison in this work will be self energy which directly contributes to the Green function and, correspondingly, to observable properties such as photoemission spectra. However, let us begin with another quantity - polarizability P, which appears at the intermediate steps of the calculation. Polarizability defines the screening effects and contributes to screened interaction W through the corresponding Dyson's equation W=V+VPW, where V is bare Coulomb interaction. The equation for W can be solved unambiguously if P satisfies certain conditions. An important one is for the "head" of polarizability

\begin{equation}
\label{head}
P^{\mathbf{k}}_{\mathbf{G}=\mathbf{G}'=0}(\nu)=\int d\mathbf{r}\int d\mathbf{r}' \frac{e^{-i\mathbf{k}\mathbf{r}}}{\sqrt{\Omega_{0}}}P(\mathbf{r},\mathbf{r}',\nu)\frac{e^{i\mathbf{k}\mathbf{r}'}}{\sqrt{\Omega_{0}}},
\end{equation}
where \textbf{r}-integration is over the whole solid, \textbf{r}'-integartion is over the volume $\Omega_{0}$ of the unit cell, \textbf{k} is momentum (point in the Brillouin zone), $\mathbf{G}$ and $\mathbf{G}'$ are the reciprocal lattice translations, and $\nu$ is bosonic frequency. The condition for the "head" of polarizability states that it should behave at small momenta as $P^{\mathbf{k}}_{\mathbf{G}=\mathbf{G}'=0}(\nu)=B(\nu)k^{2}$ in exact theory. This behavior cancels the $1/k^{2}$ divergence of the bare Coulomb interaction at small momenta. In self consistent diagrammatic approaches we normally have $P^{\mathbf{k}}_{\mathbf{G}=\mathbf{G}'=0}=A+Bk^{2}$ with A being small but not zero. The absolute value of A usually is much smaller than the absolute value of the "head" at all \textbf{k}-points with non-zero momenta. In practice, we evaluate (by fitting) the coefficients A and B and use only the $Bk^{2}$ part to proceed. The A coefficient becomes smaller when the number of diagrams increases (order by order or by using the Bethe-Salpeter equation for polarizability). To a certain degree its value also depends on numerical approximations (cutoffs) within the same diagrammatic approach. In this respect, it is important to use \textbf{k}-dependent functions in the evaluation of polarizability. If, however, we accept the local approximation for the vertex part, the "head" of the correction to polarizability becomes momentum independent with very large A coefficient. This was already discussed in Ref. \cite{arx_2105_07798} for layered Van der Waals ferromagnet CrI$_{3}$.

Figure \ref{p_nio} illustrates the above discussion using NiO as an example. The line from $\Gamma$ to $L$ point in the Brillouin zone was selected to represent the functions, but any other direction which starts at the $\Gamma$ point qualitatively is very similar. In full sc(GW+G3W2) calculation, the "head" is slightly negative at $k=0$ and there is slight improvement as compared to scGW case. This is also seen in the right window of the figure where only vertex part is shown. In "full" calculation it is positive at $\Gamma$ point which is to compensate negative value obtained in scGW. The compensation is not complete which tells us that first order correction to P is insufficient to make full polarizability perfectly correct. However, with the electronic structure being the main object for comparison in this work, the important contribution from W (and so from P) is related to the integral over the Brillouin zone which we evaluate in order to get self energy. Thus, certain average value of polarizability over the Brillouin zone represents the quantity of interest. In this respect, one can see from the graph that proper evaluation of the vertex correction to polarizability (using full momentum dependence in all diagrams) increases amplitude of polarizability but keeps its negative everywhere in the Brillouin zone except very small volume near $\mathbf{k}=0$ where the amplitude is slightly reduced. The increase of amplitude for $\mathbf{k}\neq 0$ means stronger screening which is physically reasonable because one of the inaccuracies of scGW is insufficient screening \cite{prb_57_2108}. The behavior of the "head" of polarizability obtained in both ('local' and 'locmin') single site approximations, on the other hand, is rather non-physical. Firstly, the correction to the scGW quantity is too big compared to the fully \textbf{k}-dependent evaluation. This, obviously, is a result of missing of the interference effects (mutual cancellation of off-site contributions at the intermediate steps of the evaluation of diagrams). Secondly, small momentum behavior is unacceptable. This is a direct consequence of the momentum-independent correction to polarizability inherent to single-site approximations (see right window in Fig. \ref{p_nio}). Slight momentum dependence of the corresponding curves seen in the right window is a result of phase factors in Eq. (\ref{head}). The rather nonphysical behavior near the $\Gamma$-point found in single site variants makes the above described handling of the small momentum limit (representing the "head" as $P^{\mathbf{k}}_{\mathbf{G}=\mathbf{G}'=0}=A+Bk^{2}$ and using only tensor $B$ to proceed) unstable, especially if there is an anisotropy in dielectric properties (case of LiFeAs in this study). Therefore, in single site approximations, we neglected the vertex correction to the "head" of polarizability at small momenta and we used only its "body" (i.e. matrix elements of polarizability in plane wave representation with plane wave indexes strictly corresponding to only nonzero momenta) to evaluate the contribution from the $\Gamma$-point.

In our study of $\alpha$-Ce and LiFeAs we found very similar (qualitatively) differences in polarizability evaluated with and without single-site approximation as in the case of NiO. Quantitatively, the case of NiO is the most dramatic which makes us to conclude even at this stage that spatial non-locality effects in NiO are rather strong. Let us, however, continue discussion by presenting other properties of this material.

Let us now directly compare self energies obtained with and without single site approximation. The central quantity for comparison will be vertex correction to self energy $\Delta \Sigma$ evaluated for selected band indexes and for selected high-symmetry points in the Brillouin zone. Its exact definition in symbolic language is $\Delta \Sigma=G3W2$, i.e. it represents strictly only the contribution beyond GW diagram. We have to clarify, however, that $\Delta \Sigma$ doesn't represent the difference $sc(GW+G3W2) - scGW$ because both parts (GW and G3W2) in sc(GW+G3W2) calculations include the effects of self-consistency beyond scGW and also the vertex correction to W (through polarizability). Some figures also compare full self energy $\Sigma=GW+G3W2$ where again all functions on the right hand side include vertex-self-consistency and vertex corrected W. The difference in sc(GW+G3W2) calculations abbreviated as "full", "local", or "locmin" consists only in the evaluation of vertex part (with or without single site approximation). GW part in all calculations is always performed in exactly the same way (without approximations).

\begin{figure}[t]
\begin{center}     
    \fbox{\includegraphics[width=6.5 cm]{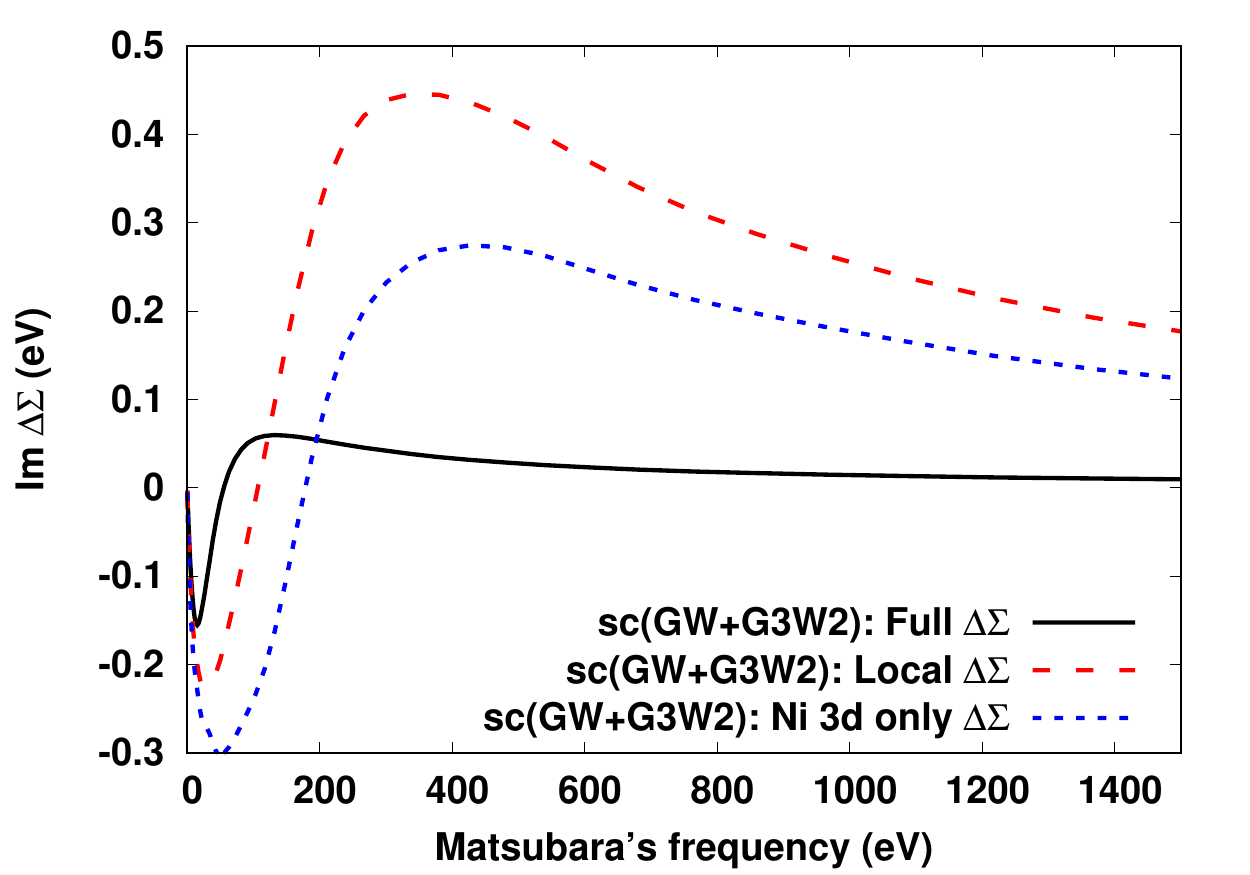}}   
    \hspace{0.02 cm}
    \fbox{\includegraphics[width=6.5 cm]{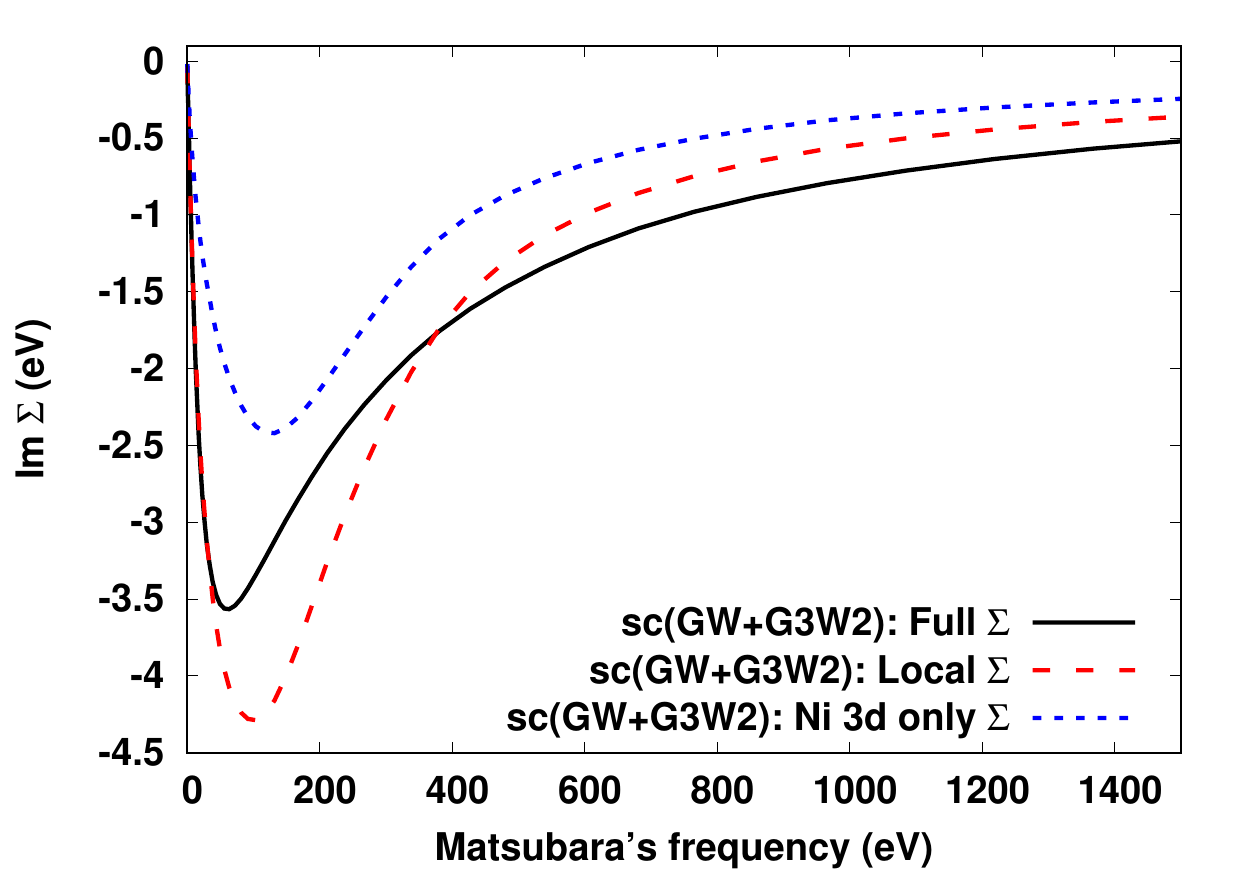}}  
    \hspace{0.02 cm}
    \caption{Imaginary part of NiO self energy as a function of Matsubara's frequency. Diagonal matrix element for the VBM band at the L point in the Brillouin zone is used for plotting. Left window shows the vertex correction, and in the right window one can see full self energy.}
    \label{sig_L_nio}
\end{center}
\end{figure}

\begin{figure}[t]
\begin{center}
    \fbox{\includegraphics[width=6.5 cm]{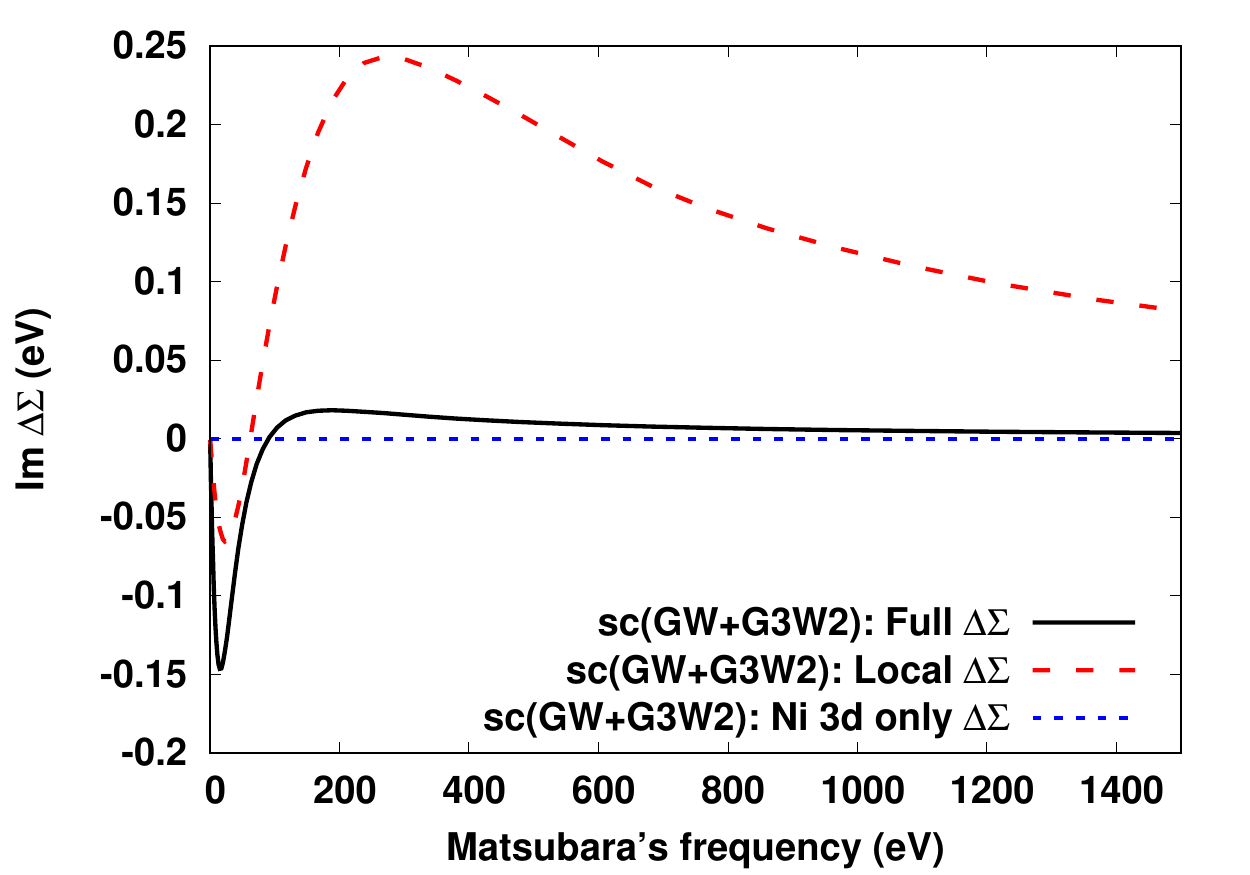}}   
    \hspace{0.02 cm}
    \fbox{\includegraphics[width=6.5 cm]{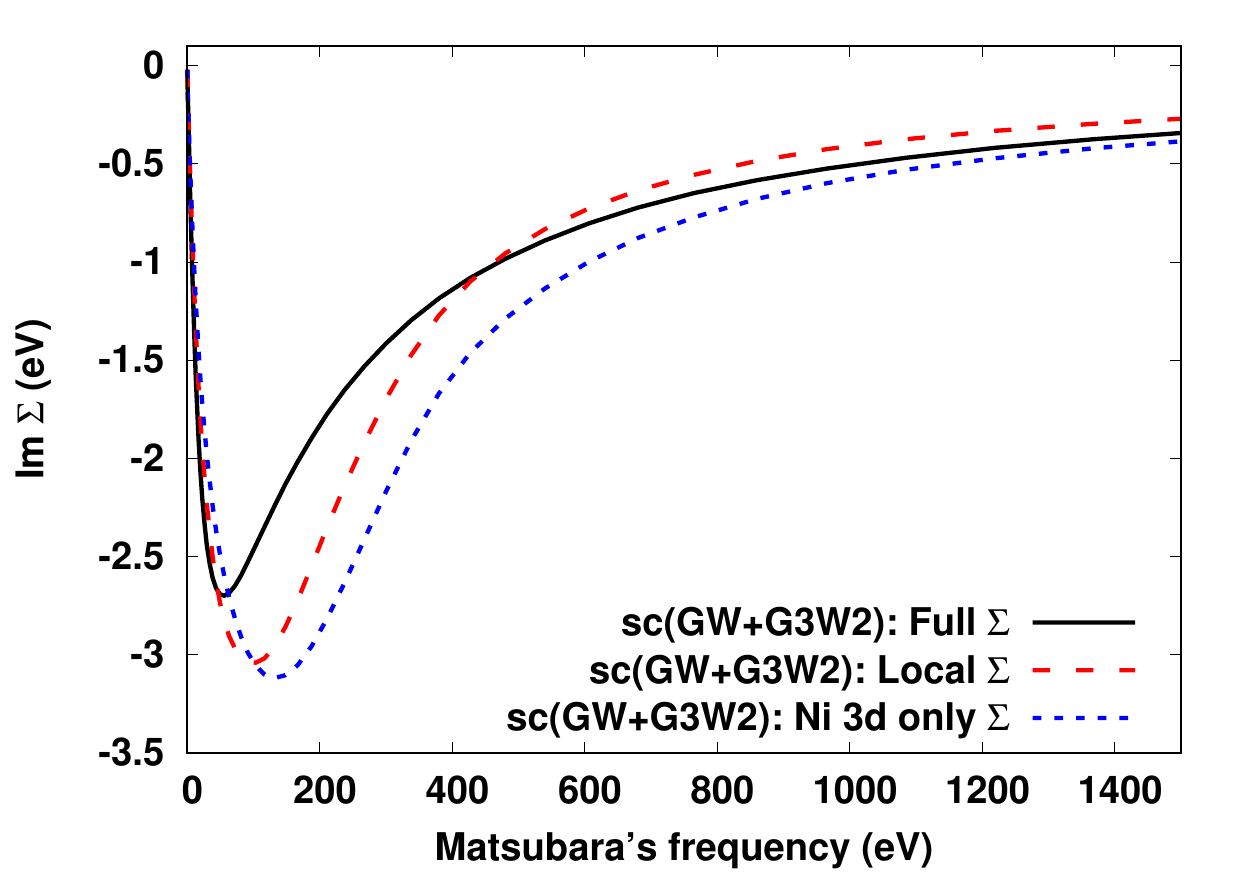}}  
    \hspace{0.02 cm}
    \caption{Imaginary part of NiO self energy as a function of Matsubara's frequency. Diagonal matrix element for the CBM band at the $\Gamma$ point in the Brillouin zone is used for plotting. Left window shows the vertex correction, and in the right window one can see full self energy.}
\label{sig_G_nio}
\end{center}
\end{figure}

\begin{figure}[t]
\begin{center}     
\includegraphics[width=9.5 cm]{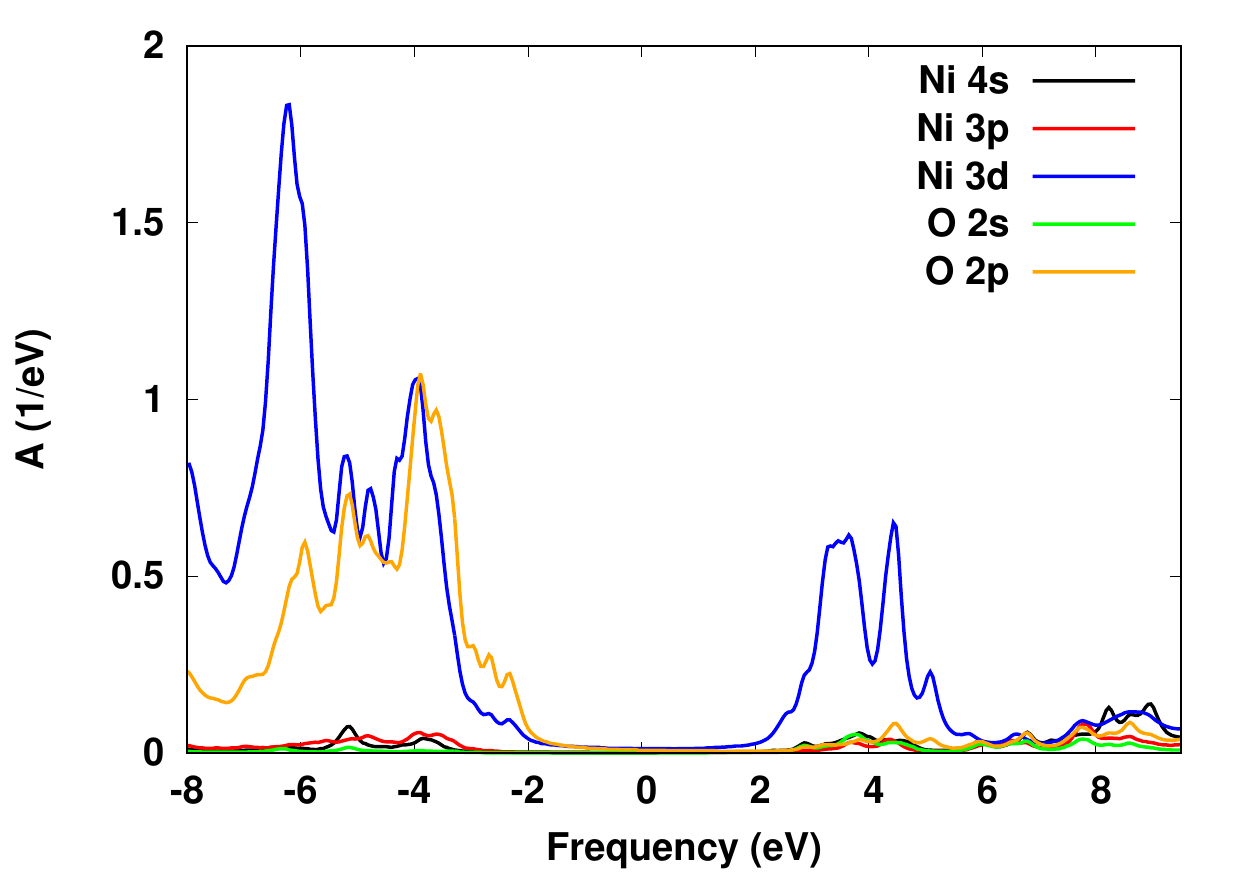}
\caption{Partial (atom and orbital resolved) spectral functions of NiO obtained with full sc(GW+G3W2) approach. Sums of spin-up and spin-down quantities are given. Analytical continuation of self energy\cite{jltp_29_179,cpc_257_107502} was used to get Green's function on the real frequency axis.}
\label{pdos_nio}
\end{center}
\end{figure}

Fig. \ref{sig_L_nio} presents the results for the valence band maximum (VBM) at $L$ point in the Brillouin zone. This band state represents a mixture of Ni 3d and O 2p orbitals. First striking difference in self energies is that correctly evaluated vertex correction (left window in Figure \ref{sig_L_nio}) is rather well "localized" in frequency space. It becomes almost negligible when Matsubara's frequency exceeds 100 eV. Similar behavior was observed before in simple semiconductors (see Fig. 18 in \cite{prb_94_155101}) and in CrI$_{3}$ (Fig. 9 in \cite{arx_2105_07798}). This effect is a result of interference (mutual cancellation) of contributions from different sites in the evaluation of higher order diagrams. The interference is missing in both single site variants and, as a result, corresponding vertex correction to self energy is rather slowly diminishing function of frequency. Second striking difference is very large (obviously non-physical) positive component of the vertex correction at frequencies larger than 100-150 eV for both single site approaches. Low frequency (negative part) of the vertex correction also demonstrates large differences in amplitude and position of the peaks. Total, i.e. GW plus vertex, self energy which is shown in the right window of Fig. \ref{sig_L_nio} represents a combined effect of both vertex corrections (to polarizability and self energy) plus the effects of self consistency. The 'local' approach shows considerable difference in amplitude and position of the principal peak and in general behavior for frequencies larger than 50 eV. But the difference between 'locmin' and 'full' functions is particularly dramatic essentially explaining huge difference in the band gaps obtained within these two approximations (see below).

Another issue related to the 'locmin' type of approximations consists in the fact that some electronic states of importance could be represented by other orbital characters (different from the ones included in 'locmin' basis set), or even by orbitals centered on other (not "correlated") sites. In order to assess vertex correction to the electronic states with orbital character not included in the 'locmin' basis set we compare vertex correction to self energy for conduction band minimum (CBM) at the $\Gamma$ point (Fig. \ref{sig_G_nio}). In fact, CBM represents a single band state which is formed from Ni 4s and O 2s. This state is almost invisible in a sea of Ni 3d band states in \textbf{k}-integrated partial spectral function shown in Fig. \ref{pdos_nio}. However, being the CBM state, it needs to be evaluated correctly. As one can see, vertex correction (left window in Fig. \ref{sig_G_nio}) in correct ('full') evaluation is also large. Its amplitude is even larger than the amplitude of correction to the VBM state. Similar to Fig. \ref{sig_L_nio}, the 'local' approximation results in slow diminishing of correction as a function of frequency and in too big positive amplitude of the correction for frequencies beyond 50 eV. Different from Fig. \ref{sig_L_nio} is that low frequency values obtained with 'local' approximation are too small as compared to the correct ones whereas in Fig. \ref{sig_L_nio} they are too big. Obviously, the 'locmin' approach fails completely for CBM state because in this case the correction is identically zero by symmetry. Total self energy shown in the right window of Fig. \ref{sig_G_nio}, again shows considerable differences between single site approaches and the 'full' one in positions and amplitudes of the peaks.

\begin{table}[t]
\caption{GW (correlation only, excluding exchange) and vertex parts of $\Psi$-functional (mRy). Vertex parts are given for full calculation (all functions are \textbf{k}-dependent), for single-site approximation, and for single-site with only one specific orbital channel.} \label{psi_1}
\begin{center}
\begin{tabular}{@{}c c c c} &NiO&Ce&LiFeAs\\
\hline\hline
$\Psi_{GW}$&-2457.9&-3074.8&-6814.2\\
$\Psi_{G3W2}$: 'full'&-13.5&-25.5&-233.9\\
$\Psi_{G3W2}$: 'local'&100.3&-29.4&-167.0\\
$\Psi_{G3W2}$: 'locmin'&34.7&-15.9&-127.7\\
\end{tabular}
\end{center}
\end{table}

One more quantity we would like to compare is the value of $\Psi$ functional, i.e. the energy associated with diagrams presented in Fig. \ref{diag_Psi}. Evaluation of $\Psi$ functional represents an important step in the calculation of electronic free energy of a material. Considering the fact that the value of functional is the result of integration over momenta and frequencies one would expect that single site approximations might be better suited for this case than for differential, i.e. momentum and frequency resolved, self energies. Table \ref{psi_1} shows the results. As one can see, the vertex part of $\Psi$ functional is about 20-100 times smaller than correlation only part of the GW functional. However, in terms of the differences between three variants of evaluation of the vertex part, NiO demonstrates a complete failure of both single site approximations. As we can see, the correct value is negative whereas both 'local' and 'locmin' variants result in positive value. This failure, therefore, can be considered as one more evidence of importance of non-local physics in the case of NiO.

The reasons for strong non-locality effects even in the diagrams beyond GW can be seen in Fig. \ref{pdos_nio} where partial (atom and orbital momentum resolved) spectral functions are presented. As one can see, upper part of the valence bands is formed predominantly by O 2p states with small admixture of Ni 3d whereas the lower conduction bands are formed mostly from Ni 3d states. Thus, most energy efficient transitions (fluctuations) which define the value of the diagrams are between oxygen and nickel sites, i.e. intrinsically non-local.

\begin{figure}[t] 
\begin{center}       
    \fbox{\includegraphics[width=6.5 cm]{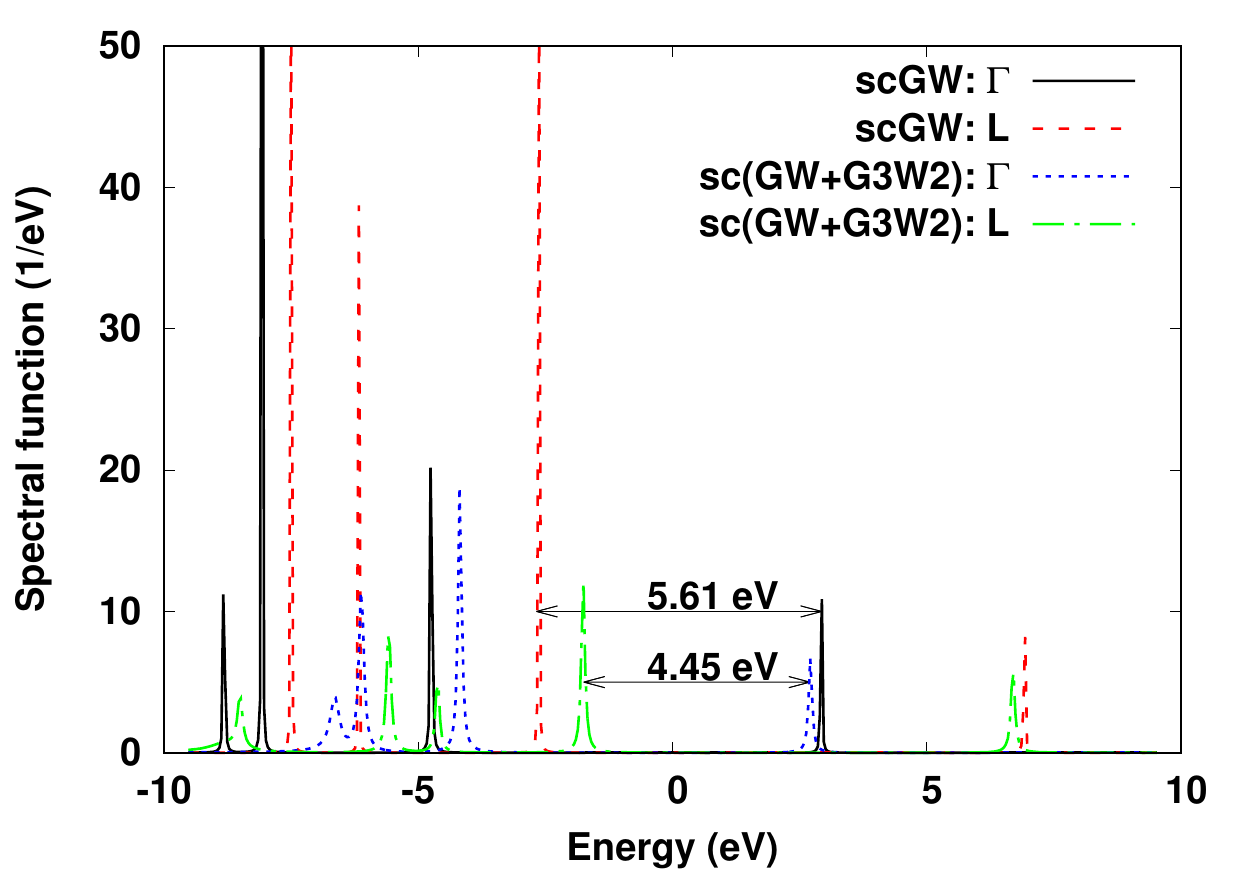}}   
    \hspace{0.02 cm}
    \fbox{\includegraphics[width=6.5 cm]{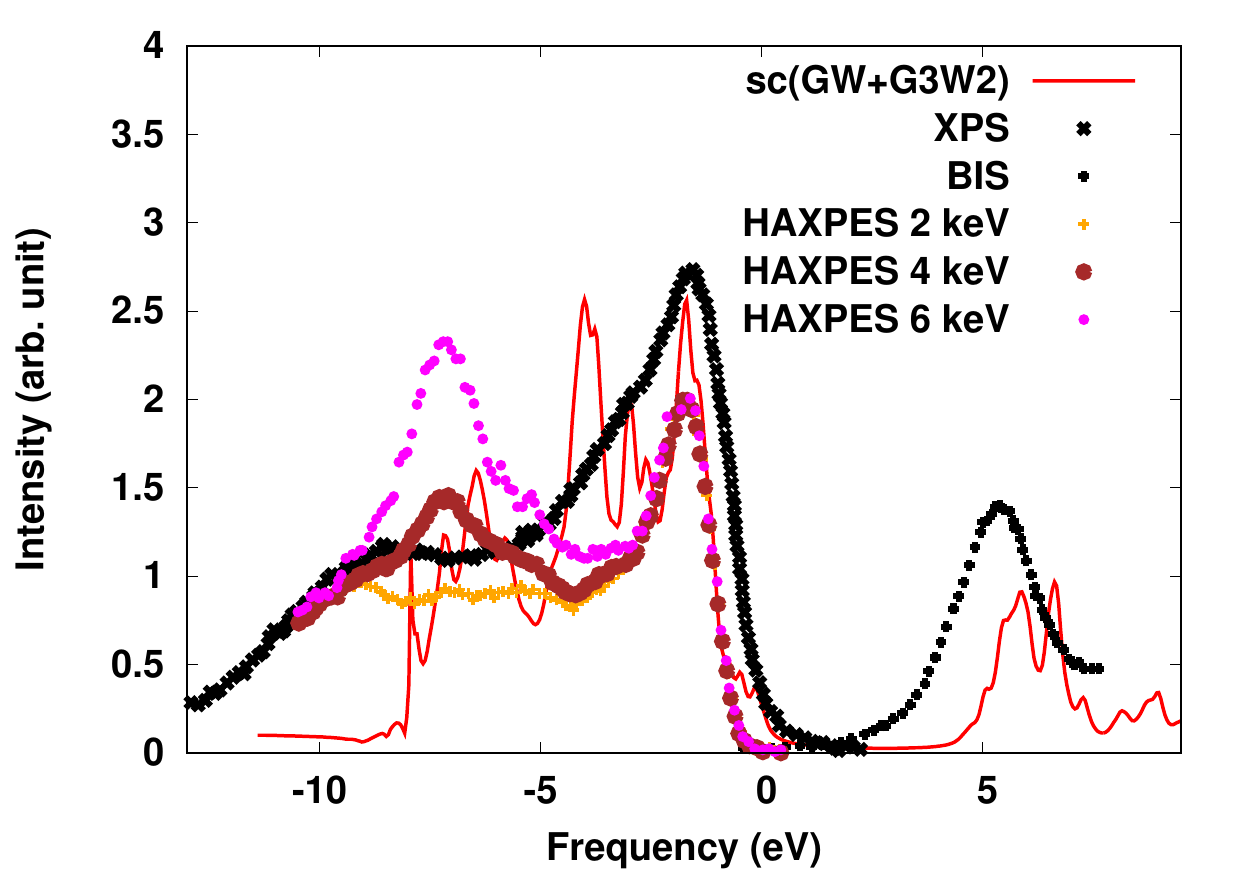}}  
    \hspace{0.02 cm} 
    \caption{Left window: Spin-up spectral function of NiO at $\Gamma$ and $L$ points in the Brillouin zone evaluated with scGW and sc(GW+G3W2) approaches. Right window: Total spectral function of NiO evaluated in sc(GW+G3W2) calculation is compared with PES+BIS \cite{prb_45_1612} and HAXPES \cite{prb_93_235138} experimental data.}
    \label{dos_nio}
\end{center}
\end{figure}

Whereas comparison with experimental data is not the principal goal of the study, it is interesting to see how sc(GW+G3W2) fares in this respect because the method is still relatively new.
Momentum-resolved spectral functions of NiO are presented in the left window of Fig. \ref{dos_nio}. Valence band maximum (VBM) for this material is at the $L$ point of the Brillouin zone and corresponds to spin-up functions, whereas the conduction band minimum (CBM) is at the $\Gamma$ point (also spin-up). Only results from scGW and full sc(GW+G3W2) calculations are shown in Fig. \ref{dos_nio}. As one can see, sc(GW+G3W2) allows to eliminate almost all discrepancy in the band gaps between scGW and experiment (4.3 eV, \cite{prl_53_2339}). In both variants of calculations with local approximation for the vertex part we found, however, an unstable behavior of iterations. During first few iterations, band gap reduced from its scGW value to approximately 2.3 eV ('local' case) and 1.5 eV ('locmin' case) and then was fluctuating around these values with amplitude of about 0.1 eV with no sign of convergence. Obviously, one can conclude that we cannot neglect non-locality effects in NiO even for diagrams beyond GW approximation. We didn't include corresponding spectral functions in Figure \ref{dos_nio} in order to avoid too many spectral lines.

\begin{figure}[t] 
\begin{center}       
    \fbox{\includegraphics[width=6.5 cm]{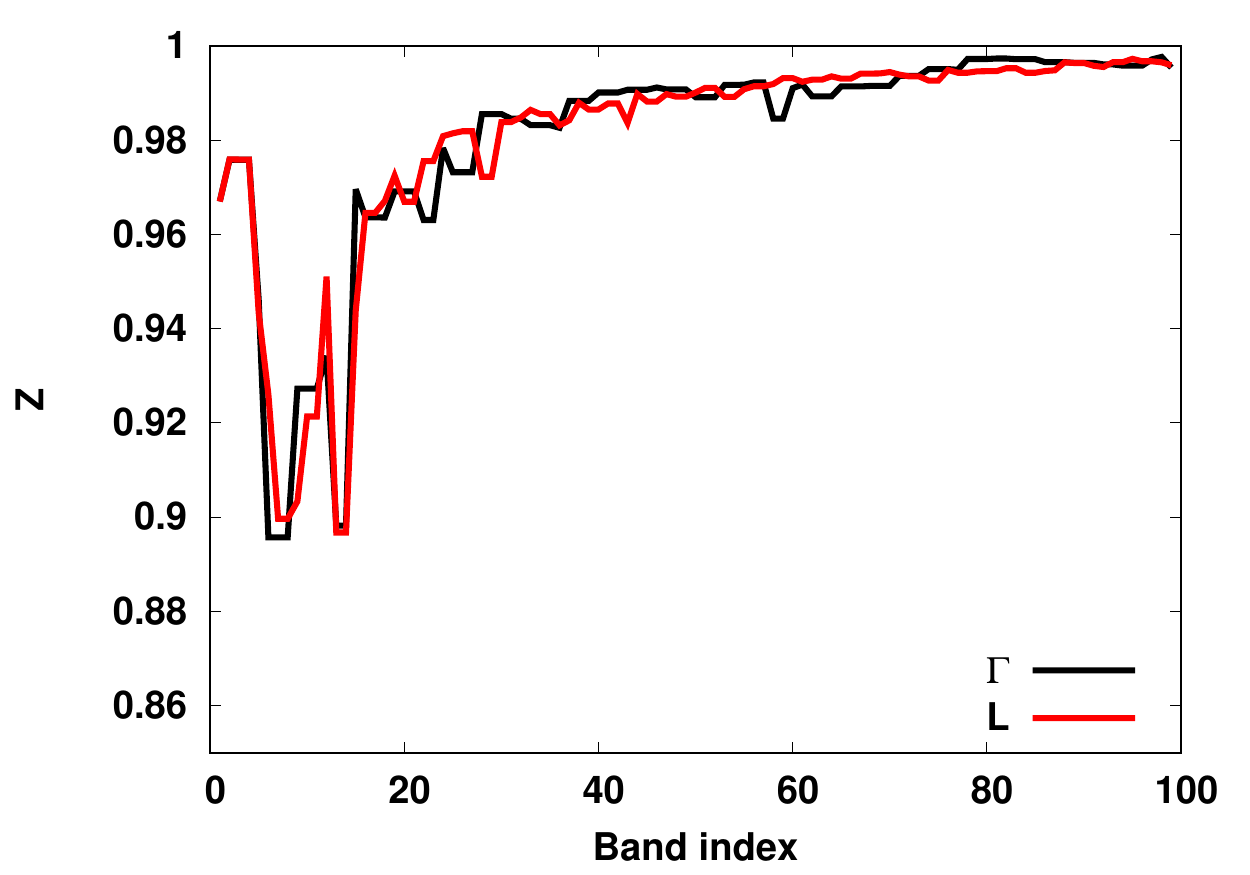}}   
    \hspace{0.02 cm}
    \fbox{\includegraphics[width=6.5 cm]{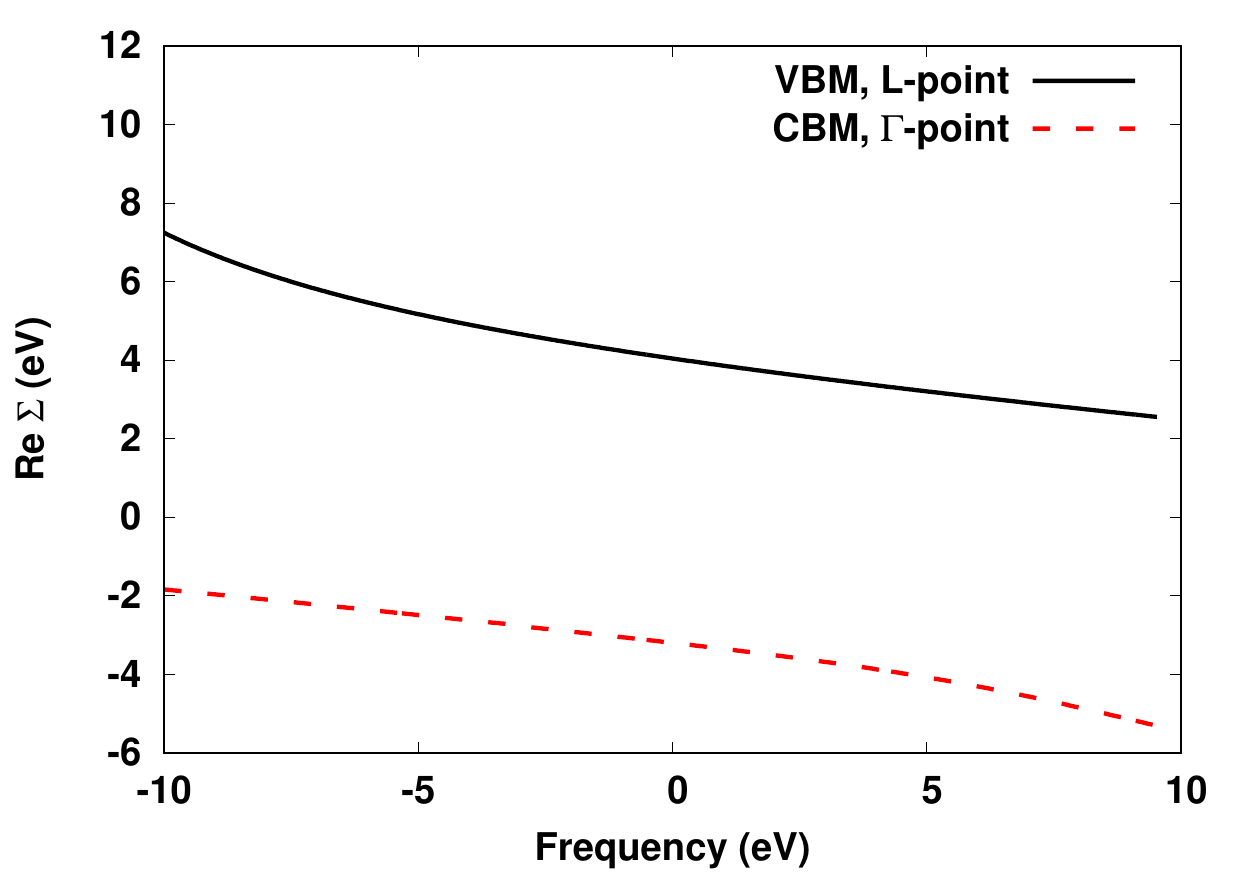}}  
    \hspace{0.02 cm} 
    \caption{Left window: Diagonal elements of renormalization factor Z of NiO evaluated in sc(GW+G3W2) calculation as a function of band index for $\Gamma$ and $L$ points in the Brillouin zone. Right window: Correlation part of self energy (NiO) as a function of real frequency from sc(GW+G3W2) calculation. Shown are real parts of diagonal matrix elements for VBM (L-point) and CBM ($\Gamma$-point).}
    \label{zre_nio}
\end{center}
\end{figure}

In the right window of Fig. \ref{dos_nio}, total spectral function of NiO as obtained in "full" sc(GW+G3W2) calculations is compared with the photoemission and inverse photoemission spectra (PES+BIS, \cite{prb_45_1612}) and with hard x-ray photoemission spectroscopy (HAXPES, \cite{prb_93_235138}) experimental data. As one can see, spectral features seen in experiments at about 7-8 eV below the gap are also decently reproduced in sc(GW+G3W2) calculations. In order to demonstrate the strength of correlation effects in ferromagnetically ordered NiO, the renormalization factor $Z_{\lambda\lambda'}^{^{-1}\mathbf{k}}=\delta_{\lambda\lambda'}-\frac{\partial\Sigma^{\mathbf{k}}_{\lambda\lambda'}(\omega)}{\partial (i\omega)}|_{\omega=0}$ has been plotted as a function of band index (Fig. \ref{zre_nio}, left window). As one can judge from the figure, minimal values of $Z$ are at the level of $0.9$ which manifests about weakness of correlation effects in magnetically ordered phase of NiO. Finally, correlation part of self energy as a function of real frequency is presented in the right window of Fig. \ref{zre_nio}. This graph is given here just for the reference purposes and can serve (for instance) to estimate (roughly) the correction of the Hartree-Fock band gap when one includes correlation effects at the level of sc(GW+G3W2).

As it was already stated above, thorough study of the properties of any specific material does not represent the main goal of this work. In respect to NiO, which traditionally is considered as strongly correlated material, and which was found in this work to be weakly correlated in its FM ordered phase, a comment has to be given. The opinion of strong correlations in NiO originates mostly from the fact that DFT calculations cannot open a gap in this material (even in its magnetically ordered phase). However, proper inclusion of nonlocal exchange-correlation effects allows one to obtain quite accurately the band gap and other important spectral features, as it was demonstrated above for magnetically ordered phase. Even application of Hartree-Fock method (i.e. only nonlocal exchange included, but no correlation) opens a gap in magnetically ordered NiO. Thus, there seems to be nothing wrong with a claim, that magnetically ordered phase of NiO (which is observed experimentally at not too high temperatures) is weakly correlated. However, in the high temperature paramagnetic phase, the situation is a lot more complicated. In author's opinion, the correct way to describe paramagnetic phase of NiO is to use spin-polarized approach but with direct modeling of thermal disorder in a large supercell. This kind of modeling was recently performed \cite{prb_97_035107} at the level of DFT+U approach with encouraging results. Whereas there are all reasons to expect that scGW or sc(GW+G3W2), when applied instead of DFT+U for a large supercell with disorder, will also result in insulating paramagnetic behavior of NiO, such calculations are totally beyond of capabilities of modern computers.

\subsection{$\alpha$-Ce}
\label{ce}

\begin{figure}[t]
\begin{center}  
    \fbox{\includegraphics[width=6.5 cm]{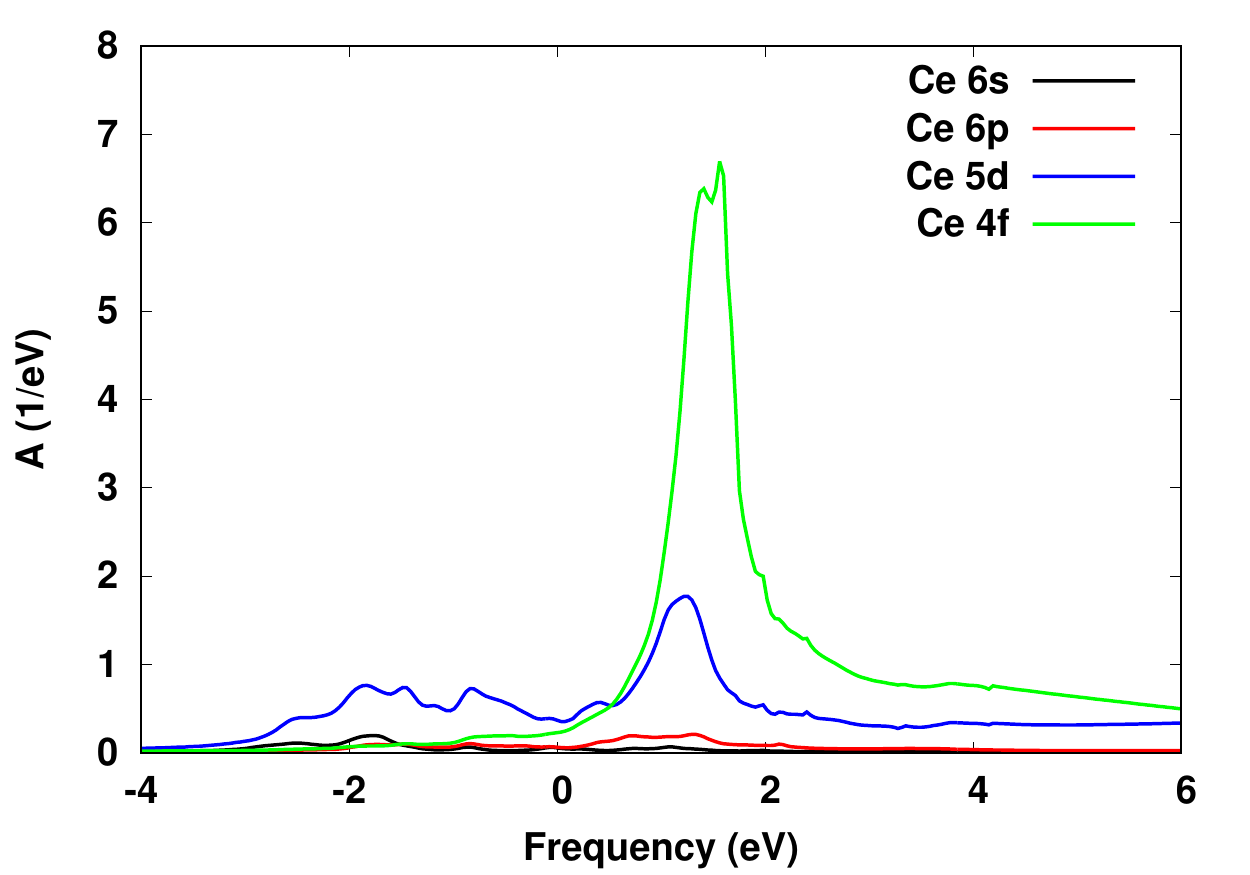}}   
    \hspace{0.02 cm}
    \fbox{\includegraphics[width=6.5 cm]{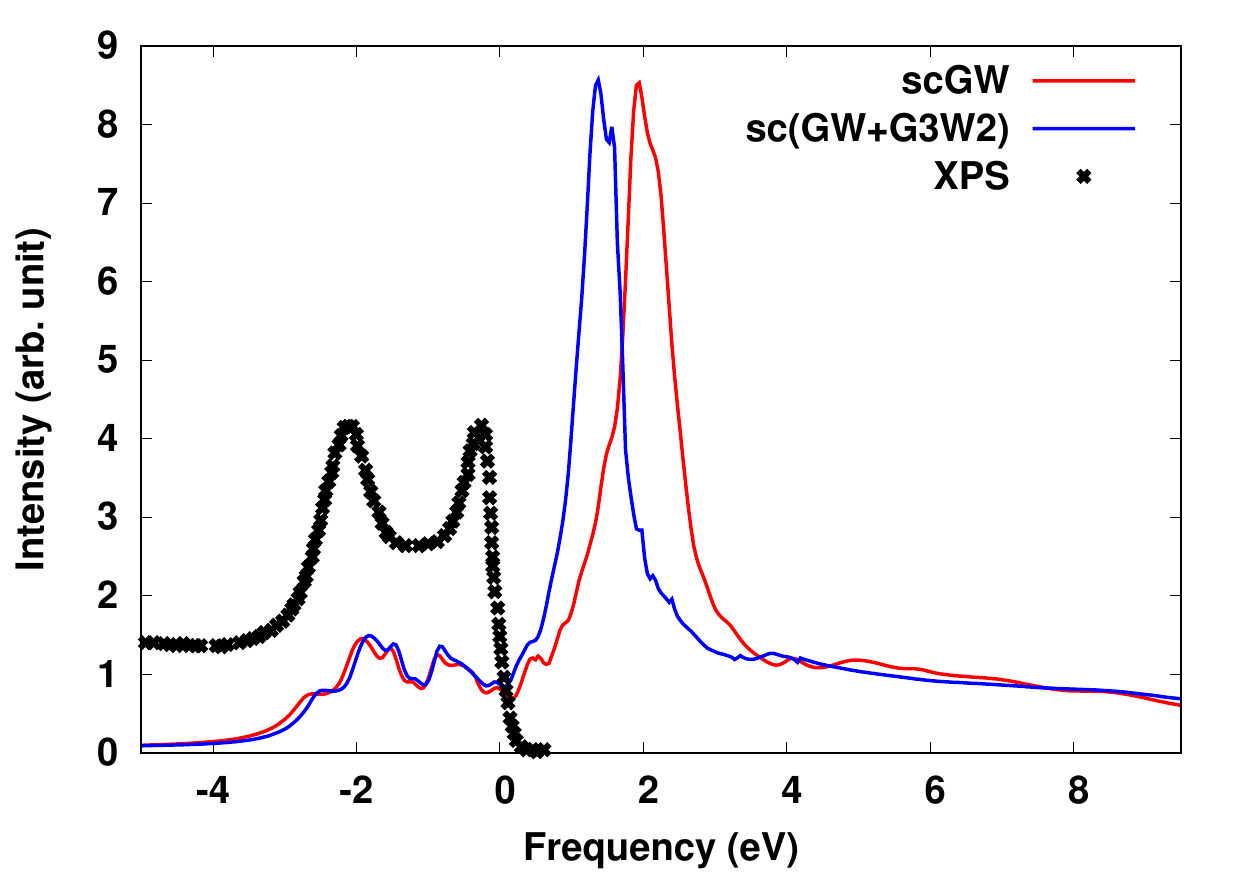}}  
    \hspace{0.02 cm}
\caption{Left: Partial (atom and orbital resolved) spectral functions of $\alpha$-Ce obtained with full sc(GW+G3W2) approach. See caption to Fig. \ref{pdos_nio} for other details. Right: Total spectral function of $\alpha$-Ce as obtained in scGW and in sc(GW+G3W2) calculations is compared with PES experimental data \cite{prb_26_7056}.}
\label{pdos_ce}
\end{center}
\end{figure}

\begin{figure}[t]
\begin{center}  
    \fbox{\includegraphics[width=6.5 cm]{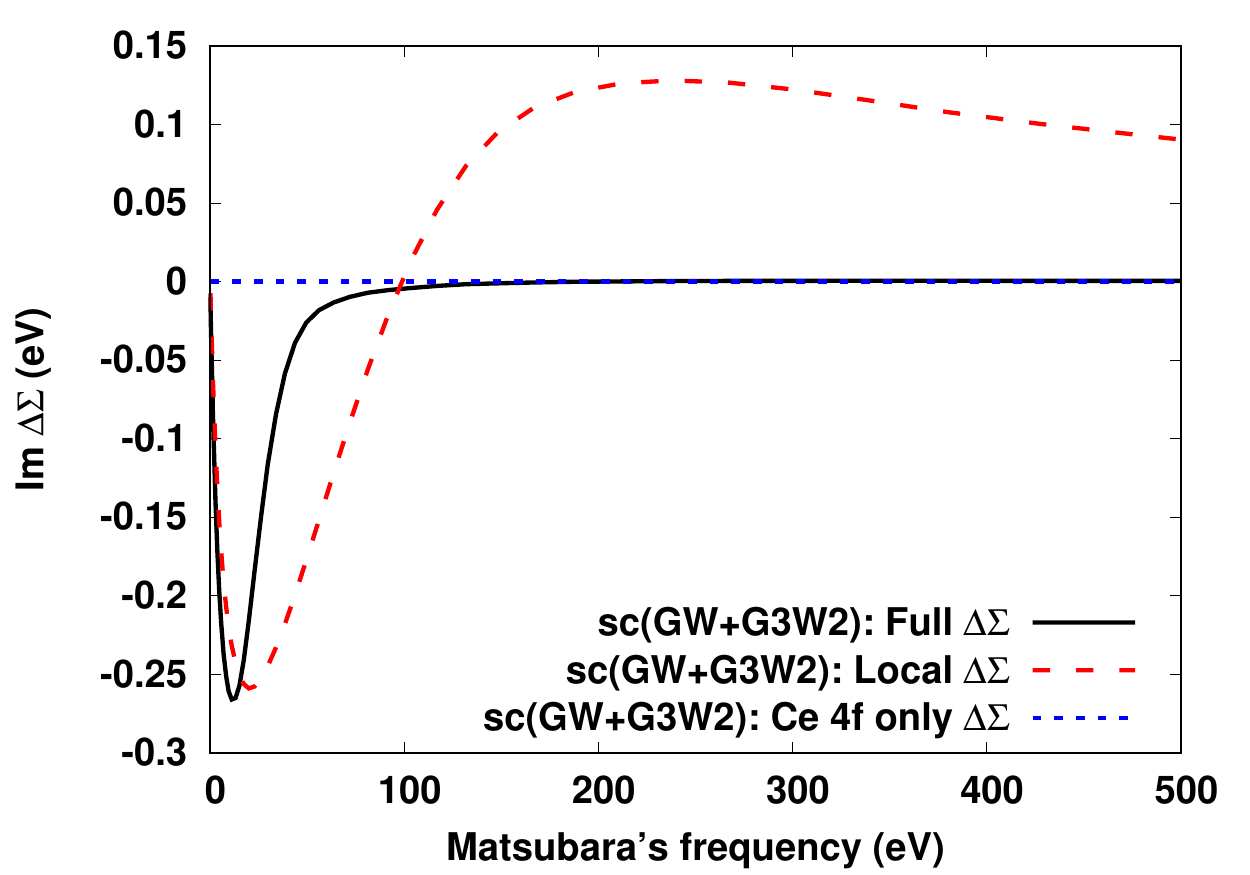}}   
    \hspace{0.02 cm}
    \fbox{\includegraphics[width=6.5 cm]{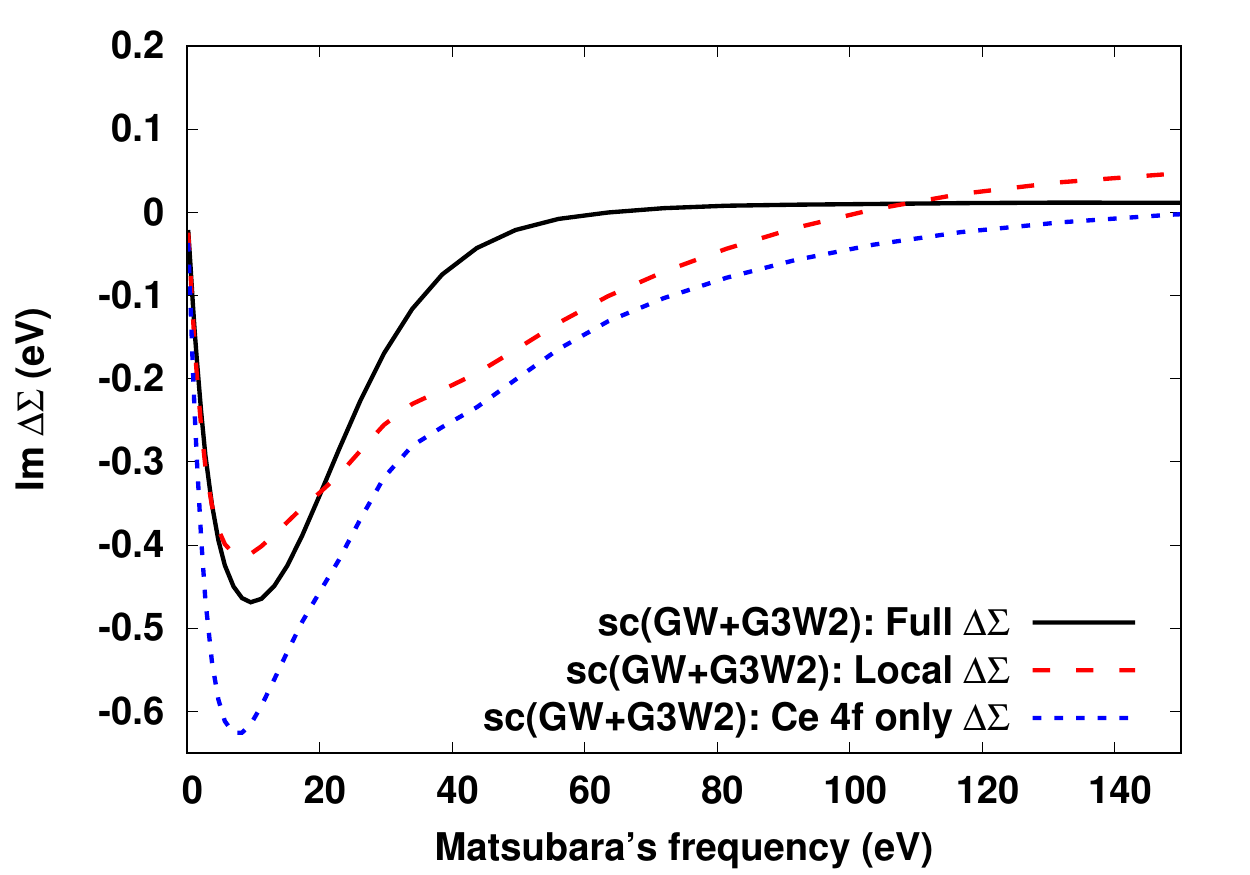}}  
    \hspace{0.02 cm}
    \caption{Imaginary part of the vertex correction to self energy for $\alpha$-Ce as a function of Matsubara's frequency. Shown are the diagonal matrix element for the spd-band at the $\Gamma$ point (left window) and for the f-band at the $X$ point (right window).}
\label{dsig_ce}
\end{center}
\end{figure}

$\alpha$-Ce belongs to the category of strongly correlated solids. It can be seen, for instance, from the right window of Fig. \ref{pdos_ce} where spectral function of this material evaluated in scGW and in sc(GW+G3W2) calculations is compared with experimental photoemission results\cite{prb_26_7056}. Principal drawback of the calculations is that they do not show essential spectral weight of f-orbitals at -2 eV and at -0.3 eV as seen in experiments. Whereas sc(GW+G3W2) approach moves the f-electron peak in the right direction (as compared to its position in scGW), it is not sufficient. Therefore, sc(GW+G3W2) does not provide enough of improvement for proper characterizing of this material. However, as it was discussed before, it is sufficient for an analysis of the quality of single site approximation. Thus, our analysis of $\alpha$-Ce should be considered simply as checking of the validity of single site approximation.

$\alpha$-Ce has one atom per primitive cell. This fact and also the fact that 4f electrons with their well localized wave functions play an important role in properties of this material makes one to expect that single site approximation should not be very bad in this case. However, looking at the partial spectral functions of $\alpha$-Ce shown in Fig. \ref{pdos_ce}, one can question the quality of the minimal 'locmin' basis set (4f only). It is clear from Fig. \ref{pdos_ce} that valence states, both occupied and unoccupied, have considerable (if not decisive) contribution of 5d orbitals with small but not negligible admixture of s and p orbitals. In order to check directly our assumption about quality of single site approximation and also our doubts about quality of minimal ('locmin') basis set, we show in Fig. \ref{dsig_ce} the vertex corrections to self energy evaluated with 'full, 'local', and 'locmin' approaches to the vertex part.

Left window of Fig. \ref{dsig_ce} shows the results for 'spd' band at $\Gamma$ point. Clearly, 'locmin' approximation (4f basis functions only) fails for this state (correction is identically zero) because of symmetry considerations. The 'local' approximation shows reasonable description of the principal negative peak (exact amplitude but slightly shifted position). However, high frequency behavior of vertex correction is still far from accurate. Vertex correction to self energy for 4f band (at $X$ point, right window in Fig. \ref{dsig_ce}) also supports the fact that the 'local' approximation works for $\alpha$-Ce slightly better than for NiO. But the 'locmin' result shows considerable deviation from the correct function in both the amplitude of the principal peak and the frequency behavior. Also, one can see that both single site approaches have some irregularities in frequency dependence (in the range 30-70 eV) which, most likely, are related to the replacement of long range functions by their local variants at intermediate steps of the diagrams evaluation.

Quick look at the value of $\Psi$ functional for $\alpha$-Ce (Table \ref{psi_1}) suggests that in the case of this material, single site approximation (but with sufficient number of local basis functions) results in relatively good value of the functional. Single site approximation with 4f orbitals as a basis results in the value of $\Psi$ functional which is almost twice less than the correct one.

\subsection{LiFeAs}
\label{lifeas}

\begin{figure}[t]
\begin{center}       
\includegraphics[width=8.5 cm]{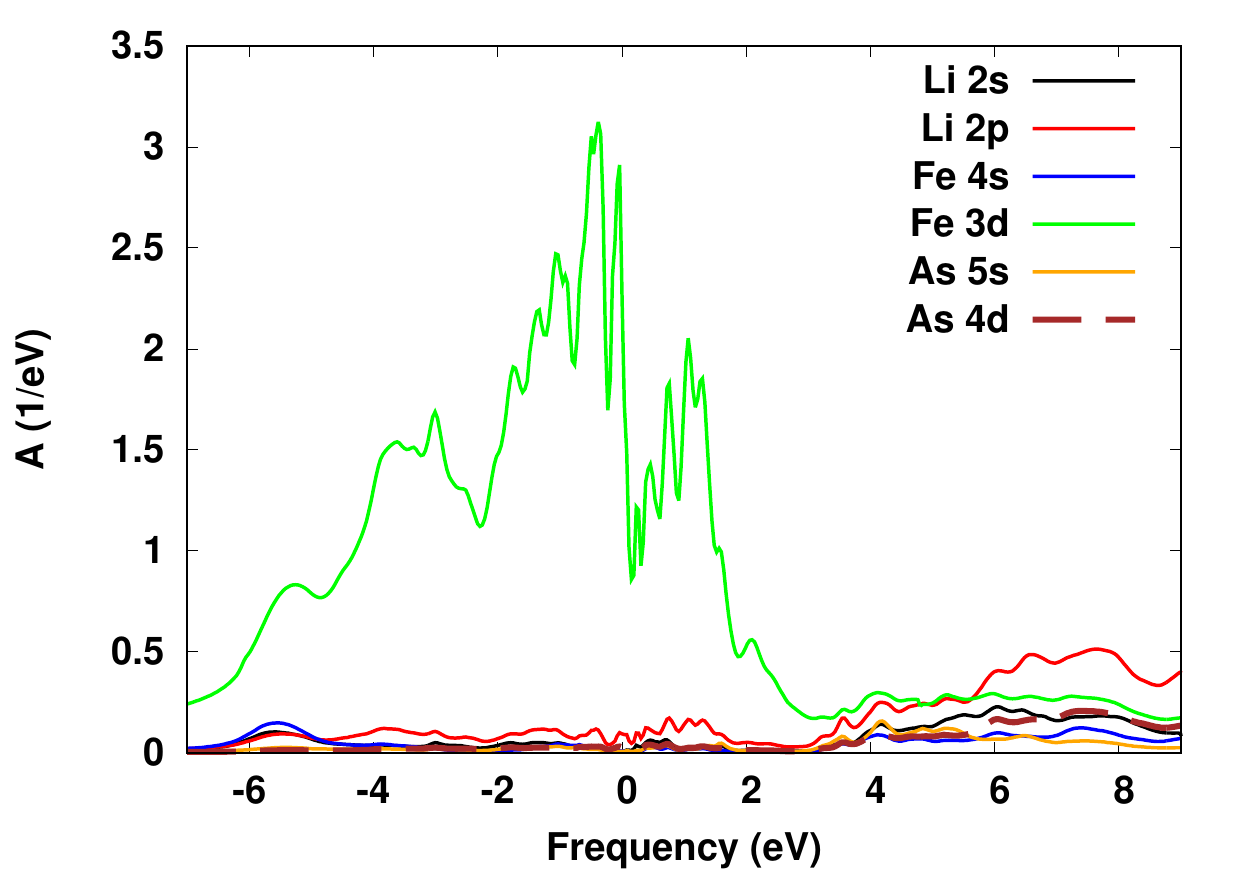}
\caption{Partial (atom and orbital resolved) spectral functions of LiFeAs obtained with full sc(GW+G3W2) approach. See caption to Fig. \ref{pdos_nio} for other details.}
\label{pdos_lifeas}
\end{center}
\end{figure}

\begin{figure}[t]
\begin{center}      
    \fbox{\includegraphics[width=6.5 cm]{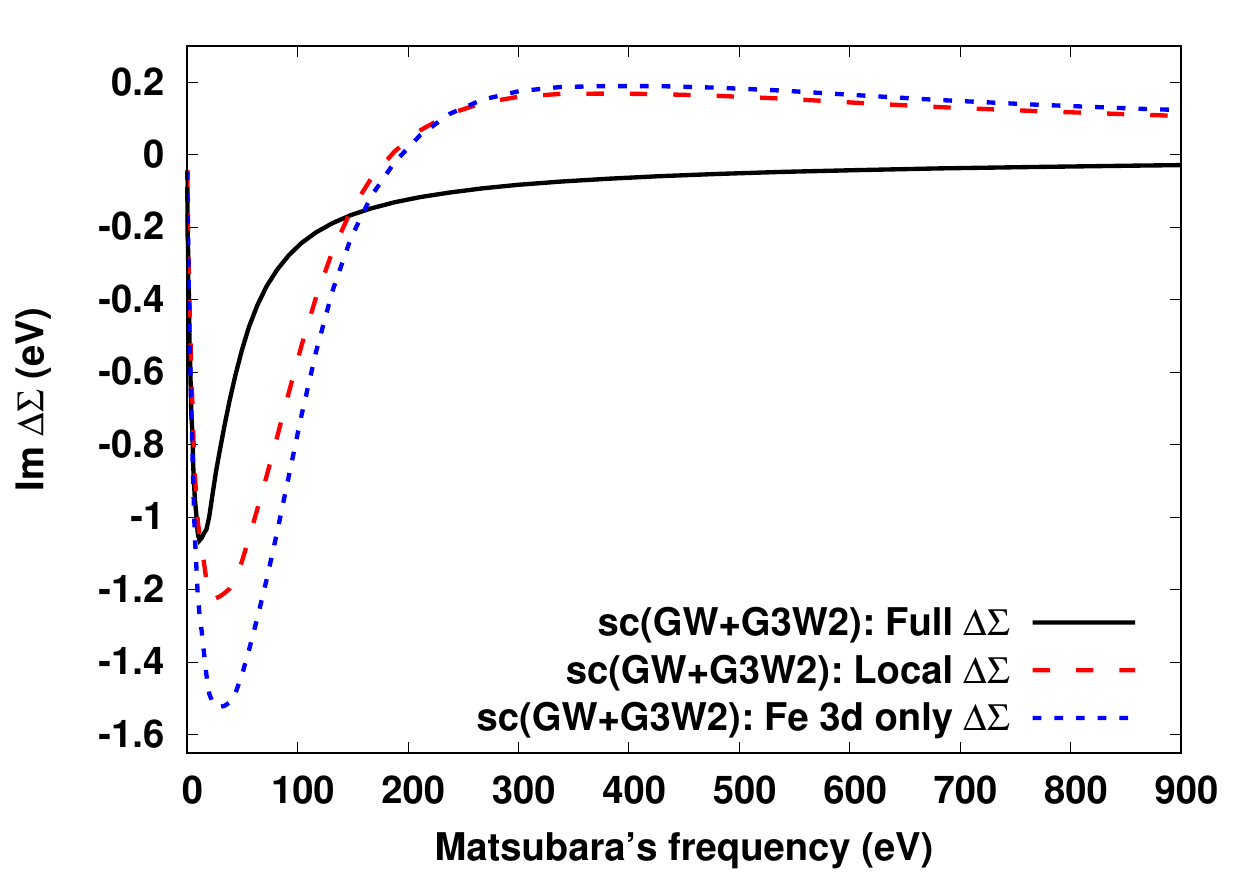}}   
    \hspace{0.02 cm}
    \fbox{\includegraphics[width=6.5 cm]{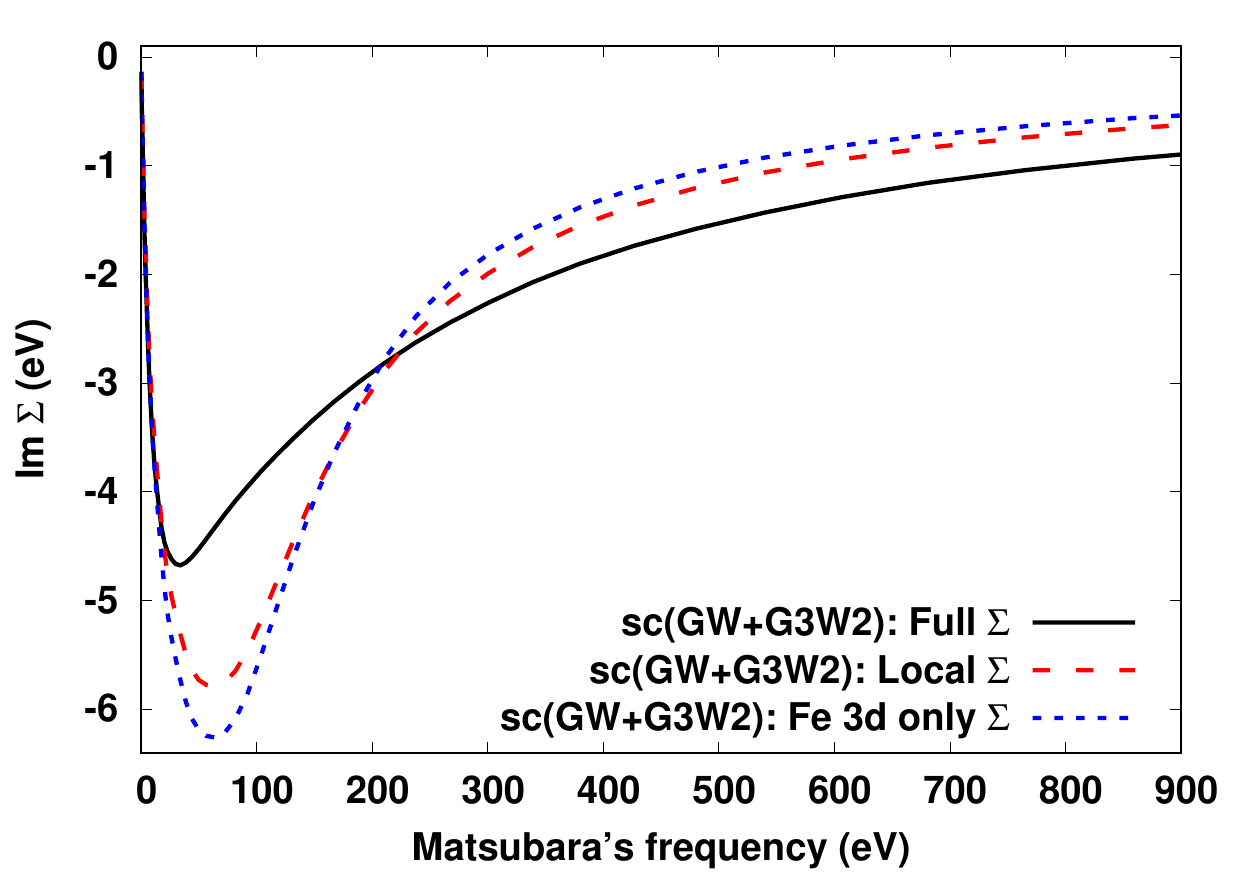}}  
    \hspace{0.02 cm}
    \caption{Imaginary part of LiFeAs self energy as a function of Matsubara's frequency. Diagonal matrix element for the VBM band at the $M$ point in the Brillouin zone is used for plotting. Left window shows the vertex correction, and in the right window one can see full self energy.}
    \label{sig_M_lifeas}
\end{center}
\end{figure}

Partial spectral function of LiFeAs presented in Fig. \ref{pdos_lifeas} demonstrates predominant Fe 3d contributions for all states near the Fermi level, both occupied and unoccupied. From this point of view, one can expect a qualitatively correct description of states near the E$_{F}$ when only Fe 3d orbitals are included in the evaluation of higher order diagrams. However, closer examination of the characters of the states near the Fermi level discovers an interesting feature that at least some of the states of importance are represented by linear combinations formed from Fe 3d orbitals but located at two different Fe sites in the primitive cell (there are two formula units in LiFeAs primitive cell). For instance, the highest valence state at the $M$ point (double degenerate) which defines the deepness of the corresponding electron pocket is formed by XZ and YZ orbitals of Fe centered at two different sites in primitive cell. Thus, one has to expect strong non-locality effects in the diagrams.

\begin{figure}[b]
\begin{center}       
\includegraphics[width=9.5 cm]{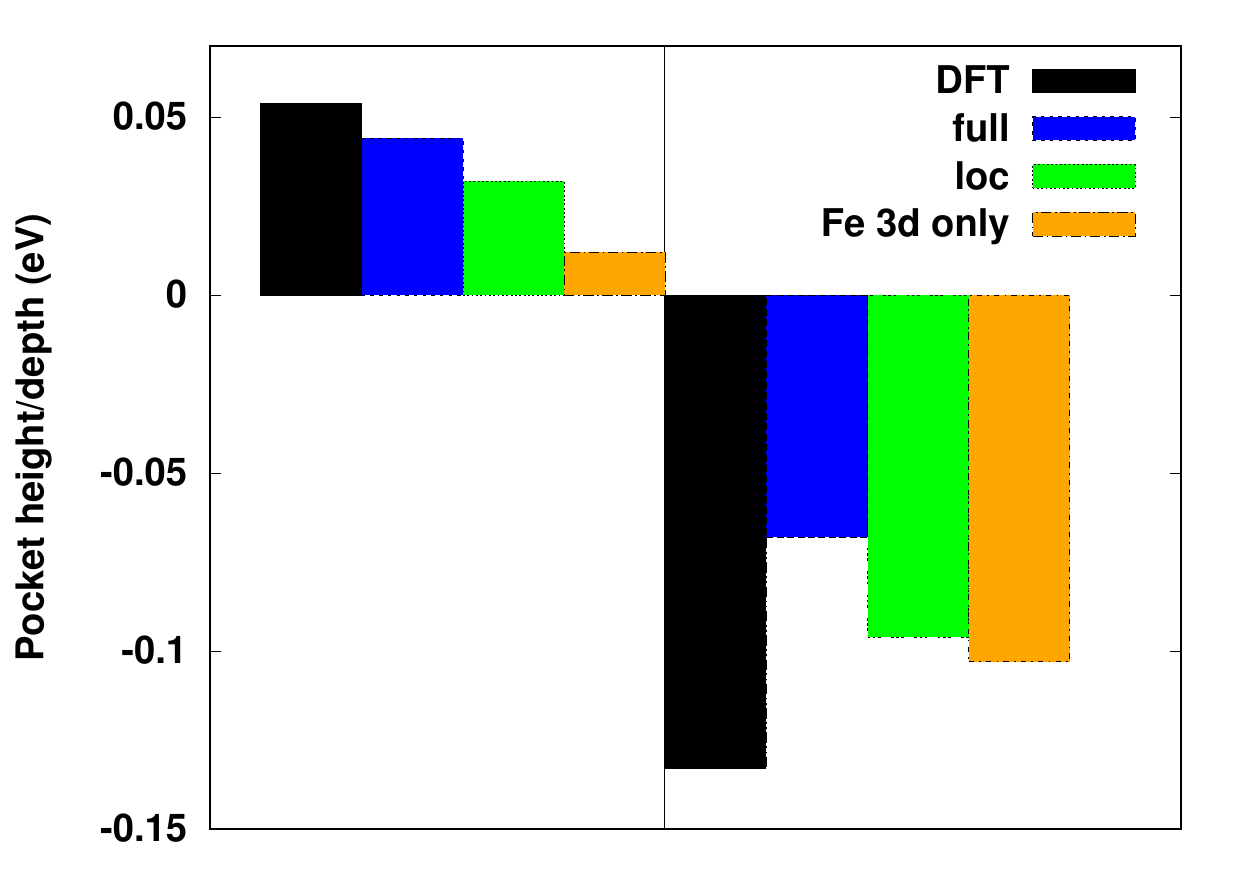}
\caption{Height of the hole pocket at $\Gamma$-point (to the left from vertical line) and depth of the electron pocket at $M$-point (to the right from the vertical line). The values were obtained from the positions of peaks of the corresponding \textbf{k}-resolved spectral functions. Presented are the results obtained in DFT and in three variants of sc(GW+G3W2) calculations.}
\label{pockets}
\end{center}
\end{figure}

Figure \ref{sig_M_lifeas} demonstrates the fact. As compared to NiO and $\alpha$-Ce, single site approximations ('local' and 'locmin') differ less between each other. But both of them deviate strongly from the 'full' approximation which supports that the spatial non-locality effects in higher order diagrams are large in LiFeAs. This conclusion is assisted by the "final" result, namely by how the vertex correction modifies scGW electronic structure near the Fermi level. Figure \ref{pockets} presents the height of the hole pocket at $\Gamma$-point (left part of the histogram) and the depth of the electron pocket at $M$ point (right part of the histogram). As one can see, both single site approximations result in too strong reduction (especially "locmin" approach) of the height of the hole pocket (we measure the effect by comparing sc(GW+G3W2) and DFT results). Also, they both result in too small reduction of the electron pocket at $M$ point. Absolute deviations in the value of $\Psi$ functional calculated with and without single site approximations are sufficiently large: 67 mRy for 'local' approximation and 106 mRy for "locmin".

\section*{Conclusions}
\label{concl}

In conclusion, we studied the effects of spatial non-locality on the higher order diagrams (beyond GW) for three materials: ferromagnetic NiO, $\alpha$-Ce, and LiFeAs. Different from all previous studies on this subject, our research was based on direct comparison of the results obtained with and without using the single site approximation. All previous studies, instead, were based on an assessment of locality of self energy without checking if the local approximation is appropriate also for the intermediate steps which have to be done in the process of evaluation of self energy.

Our results undoubtedly demonstrate that non-local effects for the higher order diagrams are important for all three materials. For NiO and LiFeAs they, in fact, are decisive. For $\alpha$-Ce, the single site approximation itself is not dramatically wrong. However, our direct comparison of the results for $\alpha$-Ce revealed another issue: using only 4f functions as a local basis set for this material is not sufficient for quantitatively correct results.

In addition to three materials studied in this work, one can also mention CrI$_{3}$ where the non-local effects were found important recently, \cite{arx_2105_07798}. Obviously, it is too early to generalize the conclusion for all materials. Probably, there are some of them where single site approximation is appropriate for certain applications. However, based on the results of this study, it seems a lot more logical to think that principal assumption of the GW+DMFT method, namely, the locality of all diagrams beyond the GW, is numerically incorrect for majority of materials.

\section*{Acknowledgments}
\label{acknow}
This work was   supported by the U.S. Department of energy, Office of Science, Basic
Energy Sciences as a part of the Computational Materials Science Program.

\appendix

\section{Specifics of the local approximation for G}\label{locappr}

In this Appendix we present precise definition of single site (local) approximation for Green's function $G$ as it is used in the main text. We will omit frequency/time argument for simplicity. We start with the representation of $G$ as a sum over the band states $\Psi^{\mathbf{k}}_{\lambda}(\mathbf{r})$:

\begin{equation}
\label{G_1}
G(\mathbf{r},\mathbf{r}')=\frac{1}{N_{\mathbf{k}}}\sum_{\mathbf{k}}\sum_{\lambda\lambda'}\Psi^{\mathbf{k}}_{\lambda}(\mathbf{r})G^{\mathbf{k}}_{\lambda\lambda'}\Psi^{^{*}\mathbf{k}}_{\lambda'}(\mathbf{r}'),
\end{equation}
where \textbf{k} is momentum, and $\lambda$ are band indexes. Because of translation invariance, argument $\mathbf{r}'$ can be restricted to be in the central unit cell whereas argument $\mathbf{r}$ can be anywhere in solid (not only in the central unit cell). Actually, expression (\ref{G_1}) is the representation of $G$ in scGW calculations (when we use "real space" to evaluate polarizability or self energy, \cite{prb_85_155129}) and in full sc(GW+G3W2) (i.e. without local approximation) calculations. Important observation is that both space arguments can be in the muffin-tin (MT) sphere or in the interstitial region.

First step in the definition of the local approximation consists in requirement that both \textbf{r} and \textbf{r}' can be only in some MT sphere, but not in the interstitial region. Mathematically it can be written with the help of the following representation for the band states (exact by construction of the LAPW+LO basis set):

\begin{equation}
\label{G_2}
\Psi^{\mathbf{k}}_{\lambda}(\mathbf{r})|_{\mathbf{t}}=\sum_{i}Z^{\mathbf{k}\lambda}_{i\mathbf{t}}\phi^{\mathbf{t}}_{i}(\mathbf{r}),
\end{equation}
with $\mathbf{t}$ being the atom in unit cell and index $i$ running over all orbitals which are used in construction of LAPW+LO basis set inside the sphere at atom $\mathbf{t}$. Specifically, index $i$ distinguishes not only between different angular momentum quantum numbers but also between the so called $\phi$, $\dot{\phi}$, and the local orbitals (LO) of the basis set LAPW+LO within the same angular momentum index. The coefficients $Z^{\mathbf{k}}_{i\mathbf{t},\lambda}$ are uniquely defined from the variational coefficients representing an expansion of band states in LAPW+LO basis set and from the augmentation procedure.

When the definition (\ref{G_2}) is used in the representation (\ref{G_1}), one obtains:

\begin{equation}
\label{G_3}
G(\mathbf{r},\mathbf{r}')|^{\mathbf{r}\in \mathbf{R}+\mathbf{t}}_{\mathbf{r}'\in \mathbf{t}'}=\sum_{ij}\phi^{\mathbf{t}}_{i}(\mathbf{r})G^{\mathbf{R}}_{i\mathbf{t};j\mathbf{t}'}\phi^{^{*}\mathbf{t}'}_{j}(\mathbf{r}'),
\end{equation}
with vectors $\mathbf{R}$ being the translation vectors (in real space). The coefficients $G^{\mathbf{R}}_{i\mathbf{t};j\mathbf{t}'}$ are, therefore, defined as the following:

\begin{equation}
\label{G_4}
G^{\mathbf{R}}_{i\mathbf{t};j\mathbf{t}'}=\frac{1}{N_{\mathbf{k}}}\sum_{\mathbf{k}}e^{i\mathbf{k}\mathbf{R}}\sum_{\lambda\lambda'}Z^{\mathbf{k}\lambda}_{i\mathbf{t}}G^{\mathbf{k}}_{\lambda\lambda'}Z^{^{*}\mathbf{k}\lambda'}_{j\mathbf{t}'}.
\end{equation}

Second step in making the "local" approximation consists in further requirement which makes things local: considering only $\mathbf{R}=0$ and $\mathbf{t}=\mathbf{t}'$. From Eq. (\ref{G_4}) it follows immediately:

\begin{equation}
\label{G_5}
G^{\mathbf{t}}_{ij}=\frac{1}{N_{\mathbf{k}}}\sum_{\mathbf{k}}\sum_{\lambda\lambda'}
Z^{\mathbf{k}\lambda}_{i\mathbf{t}}G^{\mathbf{k}}_{\lambda\lambda'}Z^{^{*}\mathbf{k}\lambda'}_{j\mathbf{t}},
\end{equation}
and, correspondingly, the single site approximation for G is following from (\ref{G_3}):

\begin{equation}
\label{G_6}
G(\mathbf{r},\mathbf{r}')|^{\mathbf{r}\in \mathbf{t}}_{\mathbf{r}'\in \mathbf{t}}=\sum_{ij}\phi^{\mathbf{t}}_{i}(\mathbf{r})G^{\mathbf{t}}_{ij}\phi^{^{*}\mathbf{t}}_{j}(\mathbf{r}').
\end{equation}

\section*{References}
\label{refer}

\bibliographystyle{elsarticle-num}
%\bibliography{Method,Actinides}

\end{document}